 \documentstyle[prd,aps,eqsecnum,floats,epsfig]{revtex}
\tighten
\begin{document}
%
 \twocolumn[
 \hsize\textwidth\columnwidth\hsize\csname@twocolumnfalse%
 \endcsname
%
\title{Meson-loop contributions to the $\rho$-$\omega$ mass splitting 
\&  $\rho$ charge radius}
%
 \vspace*{-.9cm}\hspace*{\fill}
 \parbox{5.0cm}{{\hspace*{\fill} \sf FSU-SCRI-98-130} \\
 		{\hspace*{\fill} \sf IU-NTC-98-016} \\
} \vspace*{+0.9cm}
%
\author{M.~A.~Pichowsky,$^{1,2}$ 
	Sameer~Walawalkar$^{1}$ and Simon~Capstick$^{1}$}
\address{
	$^{1}$Department of Physics 
	\& Supercomputer Computations Research Institute, \\
 	Florida State University, Tallahassee, FL 32306, USA\\
	$^{2}$Nuclear Theory Center,
	Indiana University, Bloomington, IN 47405, USA
}
\date{\today}
\maketitle
\begin{abstract}
Contributions of two-pseudoscalar and vector-pseudoscalar meson loops
to the $\rho$-$\omega$ mass splitting are evaluated in a covariant model
based on studies of the Schwinger-Dyson equations of QCD.  
The role and importance of the different time orderings of the meson loops
is analyzed and compared with those obtained within time-ordered
perturbation theory.
It is shown that each meson loop contributes less than $10\%$ of the bare
mass, and decreases as the masses of the intermediate mesons increase
beyond approximately $m_\rho / 2$.  
A mass splitting of $m_{\omega} - m_{\rho}\approx 25$~MeV is obtained from the
$\pi\pi$, $K\bar{K}$, $\omega\pi$, $\rho\pi$, $\omega\eta$, $\rho\eta$
and $K^* K$ channels.  
The model is then used to determine the effect of the two-pion
loop on the $\rho$-meson electromagnetic form factor. 
It is shown that the inclusion of pion loops increases the
$\rho$-meson charge radius by $10\%$.
\end{abstract}
]
\def\slash#1{#1 \hskip -0.5em / }  
\def\rmb#1{{\bf #1}}
\def\lpmb#1{\mbox{\boldmath $#1$}}
\def\nn{\nonumber}
\def\>{\rangle}
\def\<{\langle}
\newcommand{\Eqs}[1]{Eqs.~(\protect\ref{#1})}
\newcommand{\Eq}[1]{Eq.~(\protect\ref{#1})}
\newcommand{\Fig}[1]{Fig.~\protect\ref{#1}}
\newcommand{\Figs}[1]{Figs.~\protect\ref{#1}}
\newcommand{\Sec}[1]{Sec.~\protect\ref{#1}}
\newcommand{\Secs}[1]{Secs.~\protect\ref{#1}}
\newcommand{\Ref}[1]{Ref.~\protect\cite{#1}}
\newcommand{\Refs}[1]{Refs.~\protect\cite{#1}}
\newcommand{\Tab}[1]{Table~\protect\ref{#1}}
\renewcommand{\-}{\!-\!}
\renewcommand{\+}{\!+\!}
\newcommand{\sfrac}[2]{\mbox{$\textstyle \frac{#1}{#2}$}}
%
\sloppy
\section{Introduction}
\label{Sec:Intro}
In a fully-interacting theory, the number of constituents of a hadron
is not conserved, and the picture of a hadron as being comprised of
three quarks (baryon) or a quark and antiquark (meson) is na{\"\i}ve.
One way to go beyond a picture in which hadrons are simple bound
states of valence quarks is to consider their mixing with
multiple-hadron states.  The success of models which describe hadrons
in terms of only a few degrees of freedom, such as the
constituent-quark model, suggests that mixing with multiple-hadron
states should represent only a small correction to the predominant
valence quark structure of the hadron.

Some observables may be sensitive to contributions that arise from
mixings of hadrons with multiple-hadron states.  These observables would
provide a means to probe the structure of hadrons beyond that of the
valence quarks.  One such observable is the mass splitting between the
$\rho$ and $\omega$ mesons, which is the subject of this article.

The $\rho^0$ and $\omega$ have a similar quark substructure, differing
only in their total isospin (and hence $G$-parity).  SU(3)-flavor
violating contributions due to the different masses of the $u$, $d$ and
$s$ quarks tend to split the masses of the $\rho$ and $\omega$ mesons.  
However, these contributions alone do not
account for most of the mass splitting \cite{MT}.  Other contributions
arise from the mixing of the $\rho$ and $\omega$ mesons with
multiple-hadron states of differing $G$-parity.  An important example
of this is the shift in the mass of the $\rho$ meson due to its mixing
with the two-pion state.  As mixing between the $\omega$ meson and the
two-pion state is forbidden by $G$-parity, the mass of the $\omega$ meson 
doesn't receive such a contribution from the two-pion state.  The difference
between the $\rho$ and $\omega$ mesons due to mixing with all possible
multiple-hadron states results in an {\em observable} splitting of the
masses $m_{\rho}$ and $m_{\omega}$, which would otherwise be
degenerate (in the absence of SU(3)-flavor breaking).

In \Ref{LC}, the mass shift of the $\rho$ meson due to its mixing with
the two-pion state was investigated using an effective chiral
Lagrangian approach.  A form factor was introduced into the resulting
dispersion relations for the $\rho$-meson self energy $\Pi_{\mu
\nu}(q^2)$ to render it finite.  Several different forms of this form
factor were considered and each produced similar results.  For
example, using a dipole form factor, it was found, over a large range
of dipole widths $0.6 \leq \Lambda^2 \leq $ 4.0~GeV$^2$, that the
$\rho$-meson mass is shifted by roughly $-10$ to $-20$ MeV due to its
coupling with the two-pion state.

The purpose of the study of \Ref{LC} was to determine whether results
obtained by lattice studies of QCD for the $\rho$ meson mass are
significantly affected when the $\rho$ meson is ``unquenched''; that
is, when the $\rho$ meson is allowed to mix with two-pion states.
Such mixings with multiple-hadron states are typically neglected in
lattice studies.  It is clear from the study, which relies on very
little dynamical modeling, that the inclusion of intermediate meson
states into the vector-meson self energy $\Pi_{\mu \nu}(q^2)$ is
indeed a small effect which can be treated as a correction to the
predominant quark-antiquark structure of the vector meson.

An extensive study of the $\rho$-$\omega$ mass splitting is carried out
in \Ref{GI}. In this work the contributions of many different
two-meson intermediate states to the masses of the $\rho$ and $\omega$
are evaluated within a nonrelativistic framework, using time-ordered
perturbation theory to evaluate the loops. The three-meson vertices
and their momentum dependence are evaluated using a string-breaking
picture of the strong decays based on the flux-tube model of the
confining interaction, and quark pair creation with vacuum
($^3P_0$) quantum numbers. It is noted that the $\rho$-$\omega$
splitting is directly proportional to an OZI-violating mixing between
ground-state vector $u\bar{u}$ and $d\bar{d}$ mesons. This mixing is
due to two-meson intermediate states which can link the initial and
final mesons. It is shown in \Ref{GI} that, in the closure limit where
the energy denominators for the intermediate states are constant, the
$^3P_0$ pair-creation operator cannot link these two vector states, so
that the OZI rule is exact in this limit. The presence of a small
OZI-violating mass splitting is, therefore, interpreted as a deviation
of the sum over intermediate states from this closure limit due to
the differing properties of the physical intermediate states.

Although the authors of \Ref{GI} point out that their calculation of
this mixing is not formally complete due to the absence of pure
annihilation diagrams where the intermediate state contains only glue
(which can be written as a sum over heavier glueball states and are
likely suppressed), the sum over intermediate states is much more
extensive than that carried out here. The convergence of this sum,
although convincingly demonstrated, is rather slow in their
calculation, requiring many different meson pairs with many possible
relative angular momenta. This can be illustrated by examining the
size of individual terms in the sum; for example, the $\pi\pi$ loop
contributes $-142$ MeV to the $\rho$ mass, and the $\omega\pi$ loop
contributes $-146$ MeV (the corresponding $\rho\pi$ loop contributes
roughly three times this amount to the $\omega$ mass). It is pointed
out in \Ref{GI} that without a flux-tube overlap function and a form
factor at the pair creation vertex, both of which are necessary for a
theoretically consistent picture of the strong decays and suppress
high-momentum created pairs, this sum would not formally
converge. Note that neither of these features of the strong decay
model are essential to fitting the strong decays of mesons in this
picture, but were essential to the convergence of the loop expansion
in \Ref{GI}. A global fit to the meson decays was carried out in order
to fit the decay model (in particular the pair creation vertex
form factor).

It was concluded in \Ref{GI} that reasonable results for the
$\rho$-$\omega$ mass splitting could be obtained only when the effects
of many two-meson intermediate states are included in the loop
expansion.  Such a conclusion raises questions about the validity of
the usual picture of a meson as being well described as a bound state
of a valence quark and antiquark.  

The motivation for carrying out the
present study is to attempt to reconcile the small effects on the
$\rho$ mass due to the two-pion loop found in \Ref{LC} with the more
substantial mass shift found in \Ref{GI}, and to examine the
consequences for the convergence of the sum over intermediate states
performed in \Ref{GI}. The $\rho$-$\omega$ mass splitting is evaluated
using meson transition form factors obtained from a covariant, quantum
field theoretic model based on the Schwinger-Dyson equations of QCD.
A similar approach was employed in \Ref{MT} for the $\pi\pi$
intermediate state.  The present study extends the work of \Refs{MT,LC}
by considering the additional channels $K \bar{K}$,
$\omega \pi$, $\rho \pi$, $\omega \eta$, $\rho \eta$, and $K^{*}K$. 
(In \Ref{Hollenberg}, a non-local quark four-point interaction was used to 
study the $\rho$-$\omega$ mass splitting and it was shown that the
three-meson intermediate state $\pi\pi\pi$ produces a negligible mass
shift.  Hence, three-meson intermediate states are neglected in the
present study.)
The model employed is fit to various properties and electromagnetic form
factors of the $\pi$ and $K$, and has been shown to give reasonable
results for the decays and electromagnetic form factors of the light
vector mesons in \Ref{Hawes}. 

Although a very different framework is employed here, results that
are qualitatively and quantitatively similar to those found in
\Refs{LC} are obtained for the $\rho$-meson mass shift due to
the two-pion loop. In each of the seven channels considered here, a mass
shift of less than $10\%$ of the total mass of the $\rho$ is
observed. These mass shifts partially cancel each other when calculating
the difference $m_{\omega} - m_{\rho}$, as observed in \Ref{GI}, and we
obtain a net mass splitting of $m_{\omega} - m_{\rho} \approx $ 25~MeV.
The experimentally observed value for this mass splitting is
$m_{\omega} - m_{\rho}=12\pm 1$~MeV.  

Of course, the calculation of the $\rho$-$\omega$ mass splitting
described herein is incomplete. Other channels, such as the
two-vector-meson intermediate states, might be expected to contribute
significantly \cite{GI}.  Nonetheless, these results are consistent
with those obtained in \Ref{LC}, and indicate that the sum over
two-meson intermediate states should converge quite rapidly as the
masses of the intermediate state mesons increase, in contrast to what
is found in \Ref{GI}.

The dependence of the vector meson self energy on the masses
of the intermediate mesons is also explored in detail. It is shown that
contributions from two-meson loops with meson masses 
above $m_\rho/2$ contribute only small amounts to the mass shift, and
their contributions decrease rapidly as the masses of the intermediate
mesons are increased. This is the essential mechanism for the expected
rapid convergence of the sum over two-meson loops and is
insensitive to the details of the model meson transition form factors,
the only dynamical input employed in the present calculations.

Another observable that is sensitive to the inclusion of meson loops
is also explored in this study for the first time.  With the same
$\rho\rightarrow\pi\pi$ transition form factors used in the
calculation of the $\rho$-$\omega$ mass splitting, the contribution of
$\pi\pi$ loops to the $\rho$-meson electromagnetic form factor is
calculated. This model predicts a  $10\%$ increase in the charge radius of
the $\rho$ meson relative to that obtained in \Ref{Hawes} which considered
only the quark-antiquark structure of the vector meson.  
The inclusion of pion loops leads to a similar $10$--$15\%$ increase for
the $\pi$-meson charge radius \cite{Bender}.  
The results from the present study also indicate that of the
two-meson states considered, the two-pion state provides the
most significant correction to the vector meson charge radius.

The brief lifetime of the $\rho$ meson makes a direct measurement of
its electromagnetic charge radius impossible at present.  However, the
size of the neutral vector meson can be extracted from high-energy
data for diffractive, vector meson photoproduction on a nucleon.  It
is possible, using a semi-empirical model of diffractive
photoproduction \cite{MAP}, to relate the {\em diffractive radii} of
mesons to their charge radii.  Given the dominance of the two-pion
state this implies that the increase in the charge radius of the
$\rho$ meson due to the two-pion loop corresponds to an analogous
increase of the diffractive radius of the $\rho$ meson.  Since
$G$-parity forbids the $\omega$ meson from receiving such
contributions from the two-pion state, a measured difference in the
diffractive radii of the $\rho$ and $\omega$ mesons can provide a means
to probe the magnitude of the two-pion contribution to
$\rho$ meson observables.

The organization of this article is as follows.  In \Sec{Sec:Amps},
the general Lorentz and flavor structure of the meson transition
amplitudes, necessary for the calculation of the vector-meson self
energy and electromagnetic form factors, are derived.  These elements
are then employed in a study of the Schwinger-Dyson equation for the
vector meson propagator in \Sec{Sec:SelfEnergy}.  
The mass of the vector meson is related to the real part of the
vector-meson self energy, and the total decay width of the vector meson is
related to the imaginary part. 
These same elements are then employed in a study of the $\rho$ meson
electromagnetic form factors in \Sec{Sec:EMFF}.

In \Sec{Sec:Orders}, the contributions to the vector-meson self energy
arising for the different time orderings of the intermediate meson
propagators are investigated.  Aspects of time-ordered quantum mechanical
frameworks, such as that employed in \Ref{GI}, 
and the Lorentz-covariant Euclidean-based quantum field
theoretic framework employed here are discussed and contrasted.

In \Sec{Sec:Model}, the dynamical model used to calculate the
off-mass-shell behavior of the meson transition amplitudes is
described.  The model describes the meson transitions in terms of
loops involving nonperturbatively-dressed quarks and model
Bethe-Salpeter amplitudes that describe the quark-antiquark
substructure of the mesons.  The dressed-quark propagators and
Bethe-Salpeter amplitudes employed herein were developed and tested in
numerous studies of meson observables \cite{Hawes,Burden,NewOnes} and
in numerical studies of the quark Schwinger-Dyson equation of QCD
\cite{Frank,Maris,Wightman}.

Results for the $\rho$- and $\omega$-meson self energies are given in
\Secs{Sec:ResultsMassDep} and \ref{Sec:ResultsMassShifts}.  In
\Sec{Sec:ResultsOrders}, numerical results for each of the possible
time orderings of the two-pion-loop contribution to the self energy
are provided.  Results for $\rho$-meson electromagnetic form factor
$G_E$ and charge radius are given in \Sec{Sec:ResultsEMffs}.  The
conclusions of this study are given in \Sec{Sec:Concl}.

\section{Meson-loop contributions}
\label{Sec:Method}
In this section some of the formalism necessary to describe the
effects of two-meson loops on the vector-meson self energy and
electromagnetic form factors is presented. 

In \Sec{Sec:Amps}, the most general Lorentz-, parity- and
SU(3)-flavor-covariant amplitudes that describe the coupling of a vector
meson to two pseudoscalar mesons or a vector and a pseudoscalar meson
are presented.  The transition amplitudes are written in terms of a
coupling constant and an {\em off-mass-shell} transition form factor.
Both are defined so that the transition form factor is equal to one
when all three mesons are on shell.  In the case of the $\rho$ meson
coupling to two pions, the coupling constant associated with the
transition amplitude is related to the decay $\rho \rightarrow \pi
\pi$ and is found by experiment to be $g_{\rho \pi \pi}$ = 6.03.  The
other coupling constants considered herein as well as the
off-mass-shell transition form factors are obtained from
considerations of SU(3)-flavor invariance or the 
dynamical model described in \Sec{Sec:Model}.

In \Sec{Sec:SelfEnergy}, the Schwinger-Dyson equation for the vector
meson propagator is given in terms of the vector meson self energy
$\Pi_{\mu \nu}(q)$.  Expressions are derived that give the self energy
in terms of loop integrations involving the propagators for the two
intermediate mesons and the transition amplitudes described in
\Sec{Sec:Amps}.  The imaginary part of the vector-meson self energy is
related to the total decay width of the vector meson, and the real part
of the self energy provides a contribution to the vector meson mass.

In \Sec{Sec:EMFF}, the contribution of the two-pion loop
to the $\rho$-meson electromagnetic form factors is considered.  A
correction to the electromagnetic (EM) charge radius of the $\rho$ meson
due to pion loops is obtained, and is compared to the results obtained
from a 
study of the quark-antiquark contribution from \Ref{Hawes}.  It is
argued that since $G$-parity forbids the $\omega$ meson coupling to
two pions, and since the two-pion contribution dominates,
measurement of the difference of the $\rho$ meson and $\omega$ meson
charge radii provides a means to determine the effects of including
such intermediate meson loops.  The electromagnetic form factors are
given in terms of the same meson transition amplitudes used in
\Sec{Sec:SelfEnergy} to calculate the self energies and decay width of
the $\rho$ meson.  Hence, the calculation of the contribution of pion
loops to the $\rho$ meson EM form factor provides a further test of
the self-consistency of the present approach.

\subsection{Meson transition amplitudes}
\label{Sec:Amps}

The most general coupling of a $\rho$ meson to two pseudoscalar mesons
($\pi$ or $K$) can be written in terms of the following action:
\begin{equation}
S \!=\!\! \int\!\! d^4\!x\, d^4\!y\, d^4\!z\,\pi^i(x) \pi^j(y) \rho_{\mu}^k(z)
\Lambda^{ijk}_{\mu}(x\-z,y\-z),
\end{equation}
where $\rho^k_\mu(z)$ is the $\rho$ meson field with flavor index $k$ 
and Lorentz index $\mu$, and $\pi^i(x)$ is the pseudoscalar meson field
with flavor $i$.   The nonlocal three-point coupling is described by the
amplitude
\begin{eqnarray}
\Lambda^{ijk}_{\mu}(x\-z,y\-z) & = & 
\int\!{d^4p_1\over (2\pi)^4}{d^4p_2\over (2\pi)^4} \nn\\
& & e^{-ip_1(x-z)-ip_2(y-z)} \Lambda^{ijk}_{\mu}(p_1,p_2) 
,
\end{eqnarray}
which can be written in terms of a coupling constant $g_{ijk}$, where
$i$, $j$, and $k$ are SU(3)-flavor indices, and a form factor
$f^{VPP}(p_1,p_2)$ which is a function of the Lorentz invariants
$p_1^2$, $p_2^2$ and $q^2=(p_1+p_2)^2$.  The form factor
$f^{VPP}(p_1,p_2)$ is defined to be equal to unity, when all three
mesons are on their mass shell; that is, $f^{VPP}(p_1,p_2) = 1$ when
$p_1^2 = p_2^2 = - m_\pi^2$, $q^2 = (p_1+p_2)^2 = - m_\rho^2$.  Hence,
one may write:
\begin{equation} \Lambda^{ijk}_{\mu}(p_1,p_2) =
\frac{1}{2}(p_1\-p_2)_{\mu} 
\; g^{ijk} \; f^{VPP}(p_1,p_2).
\label{VPPvertex}
\end{equation}
In general, \Eq{VPPvertex} may also have a term which is proportional to
$(p_1+p_2)_{\mu}$.  Even if such a term were present, it could not
contribute to the self energy of an {\em on-shell} vector meson since
its contraction with a spin-1 polarization vector is zero.
For $\rho \rightarrow \pi \pi$, 
$g^{ijk} \equiv g_{\rho\pi\pi} \epsilon^{ijk}$.

From the above amplitude, one obtains the invariant Feynman amplitude for 
the decay of, say, the $\rho^0$ into two charged pions:
\begin{eqnarray} 
\lefteqn{
\< \pi^+(p_1); \pi^-(p_2) | T | \rho^0(q,\lambda) \> }
 \nn \\
&=& \varepsilon_{\mu}(q,\lambda) \left[ \Lambda^{-+3}_{\mu}(p_1,p_2) 
+ \Lambda^{+-3}_{\mu}(p_2,p_1) \right],
\\
&=& 
2 g_{\rho\pi\pi} \; p_1 \cdot \varepsilon(q,\lambda) ,
\end{eqnarray}
where $\varepsilon_{\mu}(q,\lambda)$ is the polarization vector for a 
$\rho$ meson of momentum $q$ and helicity $\lambda$. 
The resulting $\rho\rightarrow\pi\pi$ decay width is
\begin{equation} 
\Gamma_{\rho\rightarrow\pi\pi}=\frac{g^2_{\rho\pi\pi}}{4\pi} 
\frac{m_\rho}{12} \left[1-\frac{4m^2_\pi}{m^2_\rho}\right]^{3/2}
.
\label{VPP:DecayWidth}
\end{equation}
The meson transition form factor $f^{VPP}(p_1,p_2)$ does not appear in
the relation for the decay width because it is unity when all mesons
are on their mass shell.  Its value away from this point is therefore not
directly observable and must be calculated from a dynamical model.

Similarly, the coupling of an $\omega$ meson to a $\rho$ and $\pi$ meson
is described by the action: 
\begin{equation}
S \!= \! \int\!\! d^4\!x\, d^4\!y\, d^4\!z\,
\rho_{\mu}^i(x) \pi^j(y) \omega_{\nu}(z) 
\Lambda_{\mu\nu}^{ij}(x\-z,y\-z)
.
\end{equation}
Here, $\omega_{\nu}(z)$ is the $\omega$-meson field and
$\Lambda_{\mu\nu}^{ij}(x\-z,y\-z)$ is the nonlocal transition amplitude
describing the coupling of two vector mesons to a pseudoscalar
meson.  Introducing the coupling constant $g_{\omega\rho\pi}$ and form
factor $f^{VVP}(p_1,p_2)$, considerations of Lorentz covariance give the
Fourier transform of $\Lambda_{\mu\nu}^{ij}(x\-z,y\-z)$ as
\begin{equation}
\Lambda_{\mu\nu}^{ij}(p_1,p_2) =
\epsilon_{\mu\nu\alpha\beta}{p_{1\alpha}p_{2\beta} \over m_{\rho}}
\; \delta^{ij} g_{\omega\rho\pi}
f^{VVP}(p_1,p_2)
\label{VVPvertex}
\end{equation}
where $p_1$ and $p_2$ are the momenta of the $\rho$ and $\pi$ mesons,
respectively, and $f^{VVP}(p_1,p_2) = 1$ when $q^2 = (p_1+p_2)^2 =
-m^2_\omega$, $p_1^2 = -m^2_\rho$, and $p_2^2= -m^2_\pi$.  
Unlike the previous case, a lack of phase space prevents the decay $\omega
\rightarrow \rho \pi$, so that it is impossible to determine the value of
the coupling constant $g_{\omega\rho\pi}$ from experiment.
Hence, both the meson-transition form factor $f^{VVP}(p_1,p_2)$ and the
coupling constant $g_{\omega\rho\pi}$ must be determined from a model
calculation.

In \Sec{Sec:Model}, a dynamical model is described and used to determine the
transition form factors $f^{VPP}(p_1,p_2)$ and $f^{VVP}(p_1,p_2)$, 
and the coupling constant $g_{\omega\rho\pi}$.
The model has been shown to give good results for the similar, physically
accessible decays $\rho\to \pi\gamma$ and $K^*\to K\gamma$, in which a
vector meson decays into a vector and pseudoscalar \cite{Hawes}.
The value of the coupling constant $g_{\rho\pi\pi}$ = 6.03 used herein is
extracted from the experimentally observable $\rho\to \pi\pi$ decay
width using \Eq{VPP:DecayWidth}.  

\subsection{Vector-meson self energy}
\label{Sec:SelfEnergy}
The spin-1 nature of a vector meson entails the following form for the
vector-meson self energy when the vector meson is on mass shell 
($q^2 = -m_{V}^{2}$):
\begin{equation}
\Pi_{\mu\nu}(q)=\left(\delta_{\mu\nu}-{q_\mu q_\nu\over q^2}\right)\Pi(q^2)
\label{Pimunu}
\end{equation}
Away from the on-mass-shell point $q^2 = - m_{V}^2$, the vector-meson self
energy is of the general form: 
\begin{equation}
\Pi_{\mu\nu}(q)=\left(\delta_{\mu\nu}-{q_\mu q_\nu\over q^2}\right)\Pi(q^2)
+ \frac{q_{\mu} q_{\nu}}{q^2} \Theta(q^2).
\label{GenPimunu}
\end{equation}
To project out the Lorentz scalar function $\Pi(q^2)$, a 
projection operator 
$\sfrac{1}{3} {\bf T}_{\mu \nu}(q,q^2)$ where
\begin{equation}
{\bf T}_{\mu\nu}(q,-m^2) \equiv 
\delta_{\mu \nu} + q_{\mu} q_{\nu} / m^2
\label{Tdef}
.
\end{equation}
is employed.
 
In the following, the dressed vector meson propagator
$\tilde{D}_{\mu\nu}(q)={\bf T}_{\mu\nu}(q,-m_V^2) \tilde{D}(q^2)$ is
given in terms of the ``bare'' vector meson propagator $D_{\mu\nu}(q)
= {\bf T}_{\mu\nu}(q,-m_V^2) {D}_0(q^2)$ and the vector-meson self
energy $\Pi_{\mu\nu}(q) = {\bf T}_{\mu\nu}(q,-m^2_V) {\Pi}(q^2)$ which
describes the mixing of the vector meson with meson loops (and in
general, multiple-hadron intermediate states).  The ``bare''
vector-meson propagator $D_0(q^2)$ is presumed to arise from the
quark-antiquark structure of the vector meson, and does not include
the effects of multiple-meson intermediate states.  Defined in this
manner, quark confinement requires that the ``bare'' vector-meson
propagator $D_0(q^2)$ is a real-valued function of $q^2$.  The
presence of a non-zero imaginary part for $D_0(q^2)$ would entail a
loss of flux for vector-meson propagation, which is a signal for the
existence of open decay channels for the vector meson.  In our
approach, such decay channels arise from the coupling of the vector
meson with multiple-hadron states.  Mixing the vector meson with
multiple-hadron states results in a dressed, vector-meson
propagator $\tilde{D}(q^2)$ that is a complex-valued function of the
vector-meson momentum $q^2$.

The dressed, vector-meson propagator $\tilde{D}(q^2)$ is obtained by
solving the Schwinger-Dyson equation:
\begin{equation}
\tilde{D}=D_0 + D_0 \Pi \tilde{D}.
\label{DSE}
\end{equation}
This integral equation gives the dressed vector meson propagator
$\tilde{D}(q^2)$ in terms of the bare propagator $D_0(q^2)$ by 
dressing the vector meson by two-meson loops as given by $\Pi(q^2)$.  
As an integral equation, 
\Eq{DSE} necessarily implies the insertion of infinitely many
self-energy $\Pi(q^2)$ contributions into the dressed, vector-meson
propagator $\tilde{D}(q^2)$;  hence, the dressing of the vector-meson
propagator resulting from \Eq{DSE} is inherently nonperturbative.

The solution to the Schwinger-Dyson equation is
\begin{eqnarray}
\tilde{D}&=&D_0+D_0 \Pi D_0 + D_0 \Pi D_0 \Pi D_0 + \cdots
,
\nn\\
&&
=D_0 \left( \frac{1}{1 +  \Pi D_0} \right),
\end{eqnarray}
so that 
\begin{equation}
\tilde{D}^{-1}=(1 + \Pi D_0)D_0^{-1}=D_0^{-1} + \Pi
.
\label{Dinv}
\end{equation}

Near $q^2 = - m_{V}^2$, the difference between the dressed and bare
inverse propagators is given by
\begin{eqnarray}
\lefteqn{
\Pi(q^2) = \tilde{D}^{-1}(q^2) - D_0^{-1}(q^2) 
} \nn \\
&=& (1 + \delta Z) \; (q^2+m_V^2 - i m_V \Gamma)
\label{DefZ}
-(q^2+m_0^2),
\end{eqnarray}
where $m_0$ is the bare mass of the vector meson, $\Gamma$ is the
total decay width of the vector meson, and $1 + \delta Z$ is the field
renormalization for the vector meson, chosen to ensure that the
asymptotic vector meson has unit normalization.  In \Sec{Sec:Model},
the Bethe-Salpeter amplitudes, which describe the quark-antiquark
substructure of the meson, are normalized to unity.  Thus, from the
definition of the normalization $1 + \delta Z$, as given by \Eq{DefZ},
it follows that the {\em correction} to the normalization that results
from the inclusion of meson-loops is given by $\delta Z$.  For meson
loops to be considered as a small effect, $\delta Z$ must be smaller
than one.

Comparison of \Eq{DefZ} and \Eq{Dinv} gives
\begin{eqnarray}
-{\rm Im}(\Pi)&=& (1 + \delta Z) \,  m_V  \,\Gamma
\\
{\rm Re}(\Pi)&=& \delta Z q^2 + (1 + \delta Z) m_V^2 - m_0^2
\label{RePi}
.
\end{eqnarray}
\Eq{RePi} provides a means to determine the normalization $\delta Z$
from the self energy $\Pi(q^2)$:
\begin{equation}
{d\over dq^2}\left[{\rm Re}\Pi(q^2)\right]_{q^2=-m_V^2}=\delta Z,
\label{Zdef}
\end{equation}
from which it follows:
\begin{equation}
m_V^2=m_0^2+{\rm Re}\Pi(-m_V^2)
.
\end{equation}
This equation gives the mass of the vector meson $m_V$, in terms of
the bare mass $m_0$ (arising from the quark-antiquark substructure of
the vector meson) and the self energy $\Pi(q^2)$.  It can be solved
iteratively,
\begin{eqnarray}
m_V^2&=&m_0^2+{\rm Re}\Pi\left(-m_0^2-{\rm Re}\Pi(-m_V^2)\right)
\nn\\
&\approx&m_0^2+{\rm Re}\Pi(-m_0^2)
-{\rm Re}\Pi^\prime(-m_0^2){\rm Re}\Pi(-m_0^2),
\label{MassShift}
\end{eqnarray}
where the expansion assumes small $m_V^2-m_0^2$. 
This relation yields a first-order correction to the vector meson mass:
\begin{equation}
m_V^2 \approx m_0^2+{\rm Re}\Pi(-m_0^2).
\label{massShifts}
\end{equation}
From \Eq{Zdef}, one identifies ${\rm Re}\Pi^\prime(-m_V^2)$ with
$\delta Z$.  Thus, the second order correction to the vector meson
mass shift is $- \delta Z\,{\rm Re}\Pi(-m_0^2)$, which is suppressed
by a factor $\delta Z$ compared to the first-order term in
\Eq{MassShift}.  The importance of higher-order corrections can be
estimated from the momentum dependence of $\Pi(q^2)$ in the vicinity
of $m_0^2$. It is found to be sufficient (see \Sec{Sec:Results})
to include only the lowest-order term from \Eq{MassShift} to obtain a
good estimate for the dressed, vector-meson mass $m_V$.

\begin{figure}[t]
\centering{\ \mbox{\  
\epsfig{figure=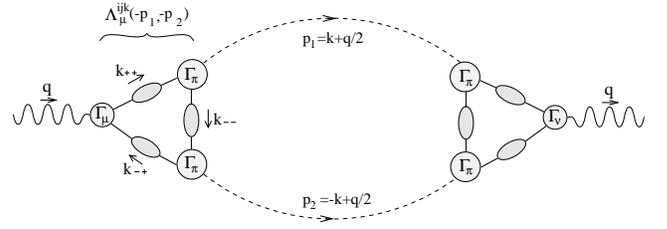,width=2.9cm,angle=-90}}}
\vspace*{0.15cm}
\caption{
Feynman diagram depicting the quark substructure of the
vector-two-pseudoscalar meson transition amplitude 
$\Lambda_{\mu}^{ijk}(p_1,p_2)$ 
(shown as a triangle-shaped quark-loop diagram) 
given by \Eq{LamVPP:Quark} 
and the intermediate pion propagators (dashed curves)
necessary for the calculation of the vector-meson self energy
$\Pi^{PP}(q^2,m_P,m_P)$.  
}
\label{Fig:Loop}
\end{figure}

As stated above, the self energy $\Pi(q^2)$ describes the dressing 
of the vector meson due to two-meson loops.  It can be described in
terms of a fluctuation of the vector meson into a multiple-meson
intermediate state.  This is given in terms of meson
transition amplitudes, the form of which was described in
\Sec{Sec:Amps}, and propagators for the intermediate mesons.
The contribution to the vector-meson self energy $\Pi(q^2)$ due to a
two-pseudoscalar-meson loop is then given by 
\begin{eqnarray}
\Pi^{kk^\prime}_{\mu\nu}(q) &=& \-2\! \int \!\!\frac{d^4\!k}{(2\pi)^4} 
\Lambda_{\mu}^{ijk}(\-p_1,\-p_2) \Delta(p_1,m_P)
\nn \\
& & \times
 \Delta(p_2,m_{P'}) \Lambda_{\nu}^{ijk^\prime }(p_1,p_2)
\nn \\
&\equiv & \delta^{kk^\prime}\Pi_{\mu\nu}(q),
\label{PimunuPP}
\end{eqnarray}
where $m_P$ and $m_{P'}$ are the masses of the pseudoscalar mesons,
$p_1 = \sfrac{1}{2} q + k$, $p_2 = \sfrac{1}{2} q - k$, 
and the meson transition amplitude $\Lambda^{ijk}_{\mu}(p_1,p_2)$, given
by \Eq{VPPvertex}, describes the transition $V \rightarrow P P$ from
a vector meson $V$ to two-pseudoscalar mesons $P$ and $P$ with momenta
$p_1$ and $p_2$, respectively.  
The propagation of the intermediate pseudoscalar mesons is given in terms 
of the following scalar-boson propagator:
\begin{equation}
\Delta(p_1,m_P)=(p_1^2+m_P^2-i\epsilon)^{-1},
\end{equation}
where $\epsilon$ is an infinitesimal positive number included to ensure the
proper boundary conditions for the decay $V \rightarrow P P$.
This process is depicted in \Fig{Fig:Loop}, in which the meson transition
amplitude $\Lambda_{\mu}^{ijk}(p_1,p_2)$ is given in terms of a quark-loop
triangle diagram.  In \Sec{Sec:Model}, the transition form factor is
calculated using the dressed-quark propagators and Bethe-Salpeter
amplitudes of a quantum field theoretic model based on studies of the
Schwinger-Dyson equations of QCD.  In impulse approximation, the leading
contribution to the meson transition amplitude 
$ \Lambda_{\mu}^{ijk}(p_1,p_2)$ is given in terms of the triangle diagram
depicted in \Fig{Fig:Loop}. 

Applying the projection operator $\sfrac{1}{3}T_{\mu\nu}(q,q^2)$ of 
\Eq{Tdef} to $\Pi_{\mu\nu}(q)$ in \Eq{PimunuPP} yields
the Lorentz-scalar vector meson self energy due to two-pseudoscalar-meson
loops
\begin{eqnarray}
\Pi^{PP}&&(q^2,m_P,m_{P'})
= \nn\\
&&-\frac{4}{3} \int\! \frac{d^4\!k}{(2\pi)^4} \Delta (p_1,m_P)
\Delta (p_2,m_{P'}) 
{\cal F}^{PP}(k,q),
\label{PiPP}
\end{eqnarray}
with
\begin{equation}
{\cal F}^{PP}(k,q)=\left(k^2- \frac{k\cdot q^2}{q^2} \right)
\big[g_{\rho\pi\pi} f^{VPP}(p_1,p_2)\big]^2
.
\label{FPP} 
\end{equation}

Similarly, the contribution of a vector-pseudoscalar-meson loop
to the vector-meson self energy is given by
\begin{eqnarray}
\Pi^{ii^\prime}_{\mu\nu}(q) &=& \int\! \frac{d^4\!k}{(2\pi)^4}
\Lambda_{\mu\sigma}^{ij}(\-p_1,\-p_2)
D_{\sigma\tau}(p_1,m_V)
\nn\ \\
& & \times
\Delta(p_2,m_P) \Lambda_{\nu\tau}^{i^\prime j }(p_1,p_2),
\label{PimunuVP}
\end{eqnarray}
where $m_V$ and $m_P$ are the masses of the intermediate vector and
pseudoscalar mesons, respectively, and the meson transition amplitude
$\Lambda_{\mu \nu}^{ij}(p_1,p_2)$ is given by \Eq{VVPvertex}.  The
propagator for the intermediate vector meson may be written as
\begin{equation}
D_{\mu \nu}(p,m_V)=\left(
\delta_{\mu \nu} + \frac{p_{\mu}p_{\nu}}{m_V^2}\right) 
\Delta(p,m_V)
.
\end{equation}
Substituting this and the relation for $\Lambda_{\mu \nu}^{ij}(p_1,p_2)$
from \Eq{VVPvertex} into \Eq{PimunuVP}, and applying the projection
operator $\sfrac{1}{3}{\bf T}_{\mu\nu}(q,q^2)$ to the result, yields 
\begin{eqnarray}
\Pi^{VP}&&(q^2,m_V,m_P)=\nn\\
&&\frac{2}{3} \int\! \frac{d^4\!k}{(2\pi)^4} \Delta (p_1,m_V)
\Delta (p_2,m_P) 
{\cal F}^{VP}(k,q),
\label{PiVP}
\end{eqnarray}
with
\begin{equation}
{\cal F}^{VP}(k,q)=
\left[\frac{(k\cdot q)^2 - k^2 q^2 }{m_{\rho}^2}\right]
\big[ g_{\omega\rho\pi} f^{VVP}(p_1,p_2)\big]^2.
\label{FVP}
\end{equation}

The flavor structure of the amplitudes in \Eqs{PimunuPP} and
(\ref{PimunuVP}) are given for the particular case of the $\rho$ meson
self energy.  It is easy to generalize the relations in \Eqs{PiPP} and
(\ref{PiVP}) to obtain the isoscalar $\omega$-meson self energy.
Assuming that the coupling constants for $V\rightarrow P P$ and $V
\rightarrow V P$ are SU(3)-flavor symmetric, one can write the self
energy of the $\rho$ meson and the $\omega$ meson in terms of the
Lorentz-scalar functions $\Pi^{PP}(q^2,m_P,m_{P'})$ and
$\Pi^{VP}(q^2,m_V,m_P)$ given by \Eqs{PiPP} and (\ref{PiVP}),
respectively, for several possible channels involving the SU(3)-octet
of vector and pseudoscalar mesons.  For simplicity, only those
two-meson loops with the lowest mass intermediate mesons are included in
the calculation of the vector-meson self energy $\Pi(q^2)$.  In
\Sec{Sec:Results}, it is shown that meson loops involving mesons with higher
masses are suppressed relative to those with lower masses and from
\Ref{Hollenberg}, contributions from three-meson loops
are expected to be very small.  For the present study, the following
channels are included: $\pi\pi$, $K\bar{K}$, $\rho\pi$, $\omega\pi$,
$\rho\eta$, $\omega\eta$, and $K^* K$.  Hence, the self energy of the
$\rho$-meson is given by
\begin{eqnarray}
\Pi(q^2)&=&
 \Pi^{PP}(q^2,m_\pi,m_\pi) + 
 \sfrac{1}{2} \Pi^{PP}(q^2,m_K,m_K)
\nn \\
& &
+ \Pi^{VP}(q^2,m_\omega,m_\pi)
+ \Pi^{VP}(q^2,m_{K^*},m_K)
\nn \\ 
& &
+ \Pi^{VP}(q^2,m_\rho,m_\eta)
, \label{PiSum:Rho}
\end{eqnarray}
and that of the $\omega$ meson is given by
\begin{eqnarray}
\Pi(q^2)&=&
0 + \sfrac{1}{2} \Pi^{PP}(q^2,m_K,m_K)
\nn \\
&&
+ 3 \Pi^{VP}(q^2,m_\rho,m_\pi)
+ \Pi^{VP}(q^2,m_{K^*},m_{K})
\nn \\ 
&&
+ \Pi^{VP}(q^2,m_\omega,m_\eta)
. \label{PiSum:Omega}
\end{eqnarray}
Once the meson transition form factors have been determined from a
dynamical model, the contributions of the self energies can be obtained
from \Eqs{PiSum:Rho} and (\ref{PiSum:Omega}) in a straight-forward manner. 

As an example, consider the calculation of the loop integral necessary
to obtain $\Pi^{PP}(q^2)$, which is illustrated in \Fig{Fig:Loop} and
is given by \Eqs{PiPP} and (\ref{FPP}), 
\begin{eqnarray}
\lefteqn{
\Pi^{PP}(q^2,m_P,m_{P'})= 
-\frac{4}{3} \int\! \frac{d^4\! k }{(2\pi)^4}
}
\nn \\
& & 
\frac{{\cal F}^{PP}(k,q)}
{[(k+q/2)^2+m_P^2-i\epsilon]
[(k-q/2)^2+m_{P'}^2-i\epsilon]}.
\label{PiPP2}
\end{eqnarray}
Introducing a Feynman parametrization
\begin{equation}
\frac{1}{a_1a_2}=\int_0^1 \frac{dx}{[(a_1-a_2)x+a_2]^2}
,
\end{equation}
the denominators of the propagators in \Eq{PiPP2} can be combined.
With a change in variables $ k\rightarrow k+\alpha q$, where
$\alpha=x-1/2$, two of the integrations can be performed trivially,
giving
\begin{eqnarray}
\lefteqn{
\Pi^{PP}(q^2,m_P,m_{P^\prime}) \hspace*{\fill}
}\nn \\
&=& \!\- \frac{4}{3} 
\int_{\-\sfrac{1}{2}}^{\sfrac{1}{2}} \!\! d\alpha \!\!\int 
\!\! \frac{d^4\! k}{(2\pi)^4}
\frac{{\cal F}^{PP}(k \- \alpha q,q)}
	{[k^{2}\-(\alpha^2\-\sfrac{1}{4}) q^2 
\+ \bar{m}^2 \+ \alpha \tilde{m}^2 \- i\epsilon]^2} ,
\nn \\
&=& \- \frac{4}{3} \frac{1}{(2\pi)^3} 
\int_{\-\sfrac{1}{2}}^{\sfrac{1}{2}} 
\!\! d\alpha \!\!\int_{\-1}^1 \!\!\! dz \sqrt{1\-z^2}
\nn \\ & & \times
\!\!\int_0^{\infty} \!\! {k^2 \, dk^2}
\frac{{\cal F}^{PP}(k\-\alpha q,q)}
{[k^2\-(\alpha^2\-\sfrac{1}{4})q^2 \+ \bar{m}^2 \+ \alpha \tilde{m}^2 
\-i\epsilon]^2},
\label{loopint}
\end{eqnarray}
where $z$ is the cosine of the angle between $k$ and $q$, defined by
$k\cdot q =\sqrt{k^2}\sqrt{q^2}\,z$, 
$\bar{m}^2 = \sfrac{1}{2}(m_{P}^2 + m_{P'}^2)$ and 
$\tilde{m}^2 = (m_{P}^2 - m_{P'}^2)$.
 An inspection of
\Eq{loopint} reveals that the denominator exhibits a zero in the limit
that $\epsilon \rightarrow 0$ for $k^2$ equal to
\begin{equation}
k^2_0 = (\alpha^2 \- \sfrac{1}{4}) q^2 - \bar{m}^2 - \alpha \tilde{m}^2
.
\end{equation}
This singular behavior leads to a simple pole in the vector-meson self
energy that is associated with the decay of the vector meson into two
pseudoscalars. 

The integral in \Eq{loopint} is evaluated numerically by first
eliminating one factor of the singularity using
\begin{equation}
\left(\frac{1}{k^2 - k_0^2}\right)^{2}
= \frac{d\;}{dk^2} \; \frac{-1}{k^2 - k_0^2}
,
\end{equation}
integrating $k^2$ by parts twice, and noting that Cauchy's theorem entails
\begin{equation}
\lim_{\epsilon \rightarrow 0^+} \frac{1}{k^2 -  k^2_0 - i\epsilon}
= \frac{{\cal P}}{k^2 - k_0^2} + i \pi\delta(k^2 - k_0^2)
, 
\end{equation}
where ${\cal P}$ denotes the principal part of the integration. 

The imaginary part of \Eq{loopint} is obtained quite easily.  To
evaluate the principal part, the $k^2$-integration domain is divided
into two regions.  The first region is from the origin ($k^2 =0$) to
twice the distance from the pole to the origin ($2 \; k_0^2 $).  The
second region is from $k^2 = 2 \; k_0^2$ to $k^2 = \infty$.  The
integrations are numerically calculated using the method of
quadratures.  In the first region of the integration over $k^2$, each
mesh point of the integration is paired with one that is equidistant
from the pole at $k^2_0$ to ensure the best cancellation of round-off
errors.  In the second region, the integration is straight-forward.
In \Sec{Sec:Amps}, it is shown that the meson transition amplitudes
$f^{VPP}(p_1,p_2)$ decrease rapidly with increasing $k^2$, providing
sufficient damping to the integrations in \Eqs{PiPP} and (\ref{PiVP})
to ensure their convergence.  Similar techniques are employed in the
calculation of the loop integral $\Pi^{VP}(q^2,m_V,m_P)$.

\subsection{Vector meson electromagnetic form factor}
\label{Sec:EMFF}
The emission or absorption of a photon by an on-mass-shell $\rho$ meson is 
described by the action:
\begin{eqnarray}
S = \int \!d^4\!x \, d^4\!y \, d^4\!z \; 
	& &\; \rho^i_{\alpha}(x) \; \rho^j_{\beta}(y) \; A_{\mu}(z)
\nn \\
& &  \times
\; \epsilon^{ij3} \; \Lambda_{\mu \alpha \beta}(x\!-z,y\!-z)
.
\end{eqnarray}
Charge conjugation and Lorentz covariance constrain the form of the
amplitude $\Lambda_{\mu \alpha \beta}(x\!-z,y\!-z)$.  Its Fourier
transform is written in terms of three Lorentz-invariant form factors
$G_i(Q^2)$ for $i=1,2,3$, which depend only on the square of the
photon momentum $Q^2 = (-q-q')^2$,
\begin{equation}
\Lambda_{\mu \alpha \beta}(q,q') = \sum_{i=1}^{3} G_{i}(Q^2) 
\;  T^{[i]}_{\mu\alpha\beta}(q,q')
\label{LamSum}
,
\end{equation}
where
\begin{eqnarray}
  T^{[1]}_{\mu \alpha \beta}(q,q^{\prime}) & = &
    (q_\mu \- q^{\prime}_{\mu}){\bf T}_{\alpha \gamma}(q,-m^2_V) 
{\bf T}_{\gamma \beta}(q^{\prime},-m^2_V) \;,\nn \\
  T^{[2]}_{\mu \alpha \beta}(q,q^{\prime}) & = &
     {\bf T}_{\mu \alpha}(q,-m^2_V) 
 {\bf T}_{\beta \delta}(q^{\prime},-m^2_V) q_\delta
\nn \\
& & 
    \- {\bf T}_{\mu \beta}(q^{\prime},-m^2_V) 
       {\bf T}_{\alpha \delta}(q,-m^2_V) 
        q^{\prime}_\delta  \:,  
\label{TiDef} \\
  T^{[3]}_{\mu \alpha \beta}(q,q^{\prime}) & = &
     \frac{q_\mu \- q^{\prime}_\mu }{2 m_V^2} 
     {\bf T}_{\alpha \gamma}(q,-m^2_V) q^{\prime}_\gamma
     {\bf T}_{\beta \delta}(q^{\prime},-m^2_V) q_\delta \;,  \nn
\end{eqnarray}
$m_V$ is the mass of the vector meson, and
\begin{equation}
{\bf T}_{\alpha \beta}(q,-m^2_V) = 
\delta_{\alpha \beta} + \frac{q_{\alpha} q_{\beta}}{m_V^2}
.
\end{equation}
One can construct a set of tensors $O^{[i]}_{\mu \alpha \beta}$ which
can be used to project out from 
$\Lambda_{\mu\alpha\beta}(q,q^{\prime})$ 
each of the form factors in \Eq{LamSum}; i.e.,
$G_i(Q^2)=O^{[i]}_{\mu\alpha\beta}\Lambda_{\mu\alpha\beta}(q,q^{\prime})$.

Alternatively, one may construct from linear combinations of the tensors
in \Eqs{TiDef} a set of tensors that transform irreducibly under spatial
rotations in a frame where there is no energy transferred to the final
$\rho$ meson, so $Q_\mu = (\vec{Q},0)$. 
The form factors associated with this set of tensors are the
electric monopole $G_E(Q^2)$, magnetic dipole $G_M(Q^2)$, and 
electric quadrupole $G_Q(Q^2)$ form factors. 
They are related to the original set of form factors by 
\begin{eqnarray}
  G_E(Q^2) & = & G_1(Q^2) + \frac{2}{3} \; \frac{Q^2}{4 m_V^2} G_Q(Q^2)\:, 
        \nn
  \\
  G_M(Q^2) & = & - G_2(Q^2)\:,  \label{FFRel}
  \\
  G_Q(Q^2) & = & G_1(Q^2) + G_2(Q^2) + (1 + \frac{Q^2}{4 m_V^2}) G_3(Q^2)\:. 
        \nn
\end{eqnarray}
Similarly, one can construct three tensors that project out these form
factors from the transition amplitude in an obvious manner, so that
$G_E(Q^2)=O^{[E]}_{\mu\alpha\beta}\Lambda_{\mu\alpha\beta}(q,q^{\prime})$.

The amplitude $\Lambda_{\mu \alpha \beta}(q,q')$ receives
contributions from many different sources.  The electromagnetic properties
of the vector meson arise from both its quark substructure as well as from
its mixings with other hadron states, such as two-pion intermediate
states.  The quark-antiquark contribution to the vector-meson EM form
factors was explored in \Ref{Hawes} using the same dynamical model
described in \Sec{Sec:Model}.  The EM form factors that arise from the
{\em quark-antiquark} substructure of the vector meson are denoted
herein as $G^{q\bar{q}}_{i}(Q^2)$.  This present study extends the
work of \Ref{Hawes} by exploring the extent to which $\pi$-meson loops
contribute to the $\rho$-meson EM form factors, denoted
$G_{i}^{\pi\pi}(Q^2)$, and $\rho$-meson charge radius.

\begin{figure}[t]
\centering{\mbox{\epsfig{figure=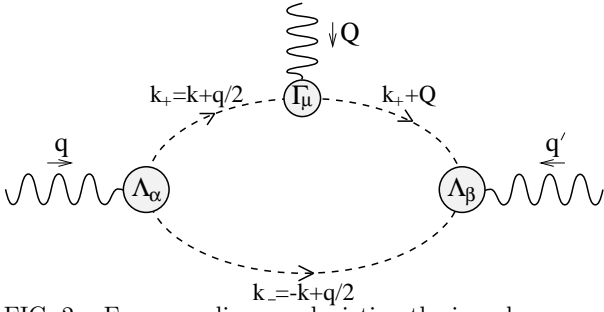,width=8.0cm}}}
\caption{ 
Feynman diagram depicting the impulse approximate two-pion loop
contribution to the $\rho$-meson electromagnetic transition amplitude
$\Lambda_{\mu\alpha\beta}(p,p')$ of \Eq{Lam2}.  } 
\vspace{1.5ex}
\label{Fig:EMpipi}
\end{figure}

In impulse approximation, the two-pion loop contribution to the
$\rho$-meson EM transition amplitude is illustrated in \Fig{Fig:EMpipi}
and is given by 
\begin{eqnarray}
\lefteqn{
\Lambda_{\mu \alpha \beta}(q,q') = 
-4i \!\! \int \!\! \frac{d^4\!k}{(2\pi)^4} 
\Delta(k_{+})  \Gamma_{\mu}(k_+\+\sfrac{1}{2}Q;Q)
}
\nn \\
& & \times 
\Delta(k_{+}\+Q) \Lambda^{3-+}_{\beta}(k_{+}\+Q,k_{-})  
\Delta(k_{-}) \Lambda^{3+-}_{\alpha}(k_{-},k_{+})
,\label{Lam2}
\end{eqnarray}
where $k_{\pm} = \sfrac{1}{2} q \pm k$, and the $\Lambda^{ijk}_{\mu}$
are as defined in \Eq{VPPvertex}.  Isospin invariance entails that the
photon couples only to the charged $\pi$ meson in the above $\pi$-loop
integration.

The $\pi$-meson EM transition amplitude, given in
terms of the usual pion EM form factor, is
\begin{equation}
\Gamma_{\mu}(k_+\+\sfrac{1}{2}Q;Q) = 
2 \left(k_+\+\sfrac{1}{2}Q\right)_{\mu} \; F_{\pi}(Q^2)
,
\label{GammaPiPi}
\end{equation}
where $k_+\+\sfrac{1}{2}Q$ is the average momentum of the incoming and
outgoing pions, and $Q=-q-q'$ is the momentum transfered to the final
vector meson by the photon.

Substituting the relation in \Eq{GammaPiPi} into \Eq{Lam2}, 
one obtains
\begin{eqnarray}
\Lambda_{\mu\alpha\beta}(q,q') &=& 8ig_{\rho\pi\pi}^2\!\!
\int \!\! \frac{d^4\!k}{(2\pi)^4}
\Delta(k_{+}) (k_{+}\+\sfrac{1}{2}Q)_{\mu} 
F_{\pi}(Q^2)
\nn \\
&\times& 
 \Delta(k_{+}\+Q) k_{\alpha}  f^{VPP}(k_{-},k_{+})  \Delta(k_{-}) 
\nn \\
&\times& 
(k\+\sfrac{1}{2}Q)_{\beta}   f^{VPP}(k_{+}\+Q,k_{-}) 
. \label{ArhorhoAmp}
\end{eqnarray}
As in the case of the vector meson self-energy $\Pi_{\mu\nu}(q)$,
the $\pi$-meson loop integration in \Eq{ArhorhoAmp} necessarily
samples momenta for which {\em both} pions are on-mass-shell when 
$q^2 = q^{\prime 2} = - m_{\rho}^2$.  
It follows that $\rho$-meson EM form factors are, in general, complex
functions of $Q^2$.

The imaginary parts of these form factors are associated with
interference between the {\em photo-stimulated} decay of the $\rho$
meson into two pions and the usual decay of the $\rho$ meson into two
pions.  The real parts of the EM form factors correspond to the
situation where the intermediate mesons remain off-mass-shell.  Hence,
they are not observed as an asymptotic state and the $\rho$ meson
remains intact.  The real parts of the EM form factors are associated
with the charge distribution of the $\rho$ meson.  Thus, only the real
parts of the $\rho$-meson electromagnetic form factors are of interest
to the present study.

As in the previous calculation of the vector-meson self energy
$\Pi_{\mu\nu}(q^2)$, a Feynman parametrization is introduced, thereby
combining the three denominators of the pion propagators in
\Eq{ArhorhoAmp} into a single denominator,
\begin{eqnarray}
\lefteqn{
\Delta(k_{+}) \Delta(k_{+}\-Q) \Delta(k_{-})
= \sfrac{1}{2} \!\! \int_{-1}^{+1} \!\!\!\! d\alpha_1 d\alpha_2
\bigg[k^2 + m_{\pi}^2 + \frac{q^2}{4}
}
\nn \\
&-&  \alpha_1 k\!\cdot\!q 
- (\alpha_1+\alpha_2)(k\! \cdot\! Q + \sfrac{1}{2} q\! \cdot\! Q
+ \sfrac{1}{2} Q^2) \bigg]^{-3}
.
\end{eqnarray}

The EM form factors of the vector meson are Lorentz invariant and can
be calculated in any frame.  Herein, the calculation is performed in
the Breit frame for which $Q_{\mu} = (0,0,Q,0)$ and
$q_{\mu}=(0,0,-\sfrac{1}{2} Q, i \sqrt{\sfrac{1}{4}Q^2 + m_{\rho}^2})$,
$k\cdot Q = x \sqrt{1 - z^2} \sqrt{k^2 Q^2}$, $k\cdot q = -\frac{1}{2}
k\cdot Q $ + $i z \sqrt{k^2(\sfrac{1}{4}Q^2 + m_{\rho}^2)}$, and the
elastic condition $2 Q\cdot q = - Q^2$ is satisfied.  Then using the
projection operators $O^{[i]}_{\mu\alpha\beta}$ discussed above and
translating the $k^2$-integration domain by
\begin{equation}
k \rightarrow \tilde k  = k + \sfrac{1}{2} \alpha_1 q 
+ \sfrac{1}{2}(\alpha_1 + \alpha_2) Q
,
\end{equation}
one finds that the two-pion contribution to the $\rho$-meson EM form
factor is given by
\begin{eqnarray}
G^{\pi\pi}_i(Q^2) &=& \frac{2i}{(2\pi)^3} \int_0^{\infty} \!\!\! k^2 dk^2
\!\!\! \int_{-1}^{+1} \!\!\!\!\! \sqrt{1\-z^2} dz dx d\alpha_1 d\alpha_2
\nn \\
& &\times \frac{ {\cal F}^i}{ [k^2 - k^2_0 - i \epsilon ]^3 },
\label{Gpipik}
\end{eqnarray}
where
\begin{eqnarray}
k_0^2 &=& - \sfrac{1}{4}(1\-\alpha_1^2)q^2 \- m_{\pi}^2 
\+ \sfrac{1}{4}(\alpha_1\+\alpha_2)(\alpha_2\+1)Q^2 , \\
{\cal F}^{i} &=& O^{[i]}_{\mu \alpha \beta} 
\big[k+\sfrac{1}{2}(\alpha_1+1)q + \sfrac{1}{2}(\alpha_1+\alpha_2+1)Q
     \big]_{\mu}
k'_{\alpha} k^{''}_{\beta} 
\nn \\
& & \times F_{\pi}(Q^2) g_{\rho\pi\pi}^2 
f^{VPP}(\tilde{k}_{-},\tilde{k}_{+}) 
f^{VPP}(\tilde{k}_{+}\+Q,\tilde{k}_{-}) 
\\
k^{\prime} &=& k \+ \sfrac{1}{2} \alpha_1 q \+ 
\sfrac{1}{2}(\alpha_1\+\alpha_2)Q, 
\\ 
k^{\prime\prime} &=& k^{\prime} + \sfrac{1}{2} Q
\\
\tilde{k}_{\pm} &=& \pm 
\left[  
k + \sfrac{1}{2} \alpha_1 q + \sfrac{1}{2}(\alpha_1 + \alpha_2) Q
\right]+ \sfrac{1}{2} q
.
\end{eqnarray}

Performing the $k$ integration by parts twice, \Eq{Gpipik} yields
\begin{eqnarray}
\lefteqn{
G^{\pi\pi}_i(Q^2) = \frac{i}{(2\pi)^3} 
\int_{-1}^{+1} \!\!\!\!
\sqrt{1\- z^2} dz \, dx \, d\alpha_1 \, d\alpha_2 
}
\nn\ \\
& & 
\times \bigg( \int_0^{\infty} \!\!\!\! dk^2
\frac{1}{ [k^2 \- k^2_0 \- i \epsilon ] }
\big[\frac{\partial\;}{\partial k^2}\big]^2  \; k^2 {\cal F}^i 
- \frac{ {\cal F}^i }{ k^2_0 }\big|_{k^2 = 0}  \bigg).
\label{Gpipik2}
\end{eqnarray}
The {\em real} part of this amplitude is of interest, which is
obtained from the principal parts of the integrations.  As in the
calculation of the vector-meson self energy $\Pi_{\mu\nu}(q)$
described in previous sections, calculation of the real part of such
amplitude requires a knowledge of the off-mass-shell transition form
factor $f^{VPP}(p_1,p_2)$ appearing in ${\cal F}^i$.  
This is the same form factor that appears in \Eq{FPP} for the
calculation of the vector-meson self energy $\Pi^{PP}(q^2)$.
The determination of this form factor is carried out in \Sec{Sec:Model},
and the results of numerical calculations of the 
$\rho$-meson EM form factors will be postponed until \Sec{Sec:Results}.
However, it is important to note that the fact that $f^{VPP}(p_1,p_2)$
appears in the calculations of both $\Pi_{\mu \nu}(q)$ and $G_E(Q^2)$
provides a self-consistency check of the dynamical model for
$f^{VPP}(p_1,p_2)$.   
For a dynamical model to be acceptable, it must lead to an off-mass-shell
transition form factor $f^{VPP}(p_1,p_2)$ that when substituted into
\Eqs{PiPP} and (\ref{Gpipik}) gives reasonable results for both the
$\rho$-$\omega$ mass splitting and the $\rho$-meson electromagnetic form
factor.  This issue is discussed further in \Sec{Sec:Results}.

Since the $\rho$-meson EM form factor receives contributions from its
quark substructure as well as from meson loops, the $\rho$ meson
electric-monopole form factor is written as the sum:
\begin{equation}
G_{E}(Q^2)  = \frac{G^{q\bar{q}}_{E}(Q^2)  + G^{\pi\pi}_{E}(Q^2)}
{G^{q\bar{q}}_{E}(0)  + G^{\pi\pi}_{E}(0)}
.
\label{Ge}
\end{equation}
Here, the factors in the denominator ensure the unit charge of the $\rho$
meson $G_E(0) = 1$. 

In \Ref{Hawes}, it was noted that the quark-photon vertex obeys the Ward 
identity which results from the $U(1)$-gauge invariance of QED.  When
quark-photon vertex satisfies this identity and the Bethe-Salpeter
amplitudes for the $\rho$ meson are normalized in the canonical
manner, the charge conservation condition [$G^{\bar{q}q}_E(0) = 1$]
for the $\rho$ meson is guaranteed.  A similar Ward identity
constrains the behavior of the pion-photon vertex, 
\begin{equation}
\lim_{Q\rightarrow 0} \Gamma_{\mu}(p,p-Q) = 
\frac{d\;}{d p_{\mu} } \, \Delta^{-1}(p)
. \label{WardId}
\end{equation}
This identity is satisfied by the form introduced in \Eq{GammaPiPi}.
It is straight-forward to show that this identity provides a relation
between the magnitude of the form factor $G_E^{\pi\pi}(0)$ and the
derivative of the self energy $\Pi^{PP}(-m_{\rho}^2,m_{\pi},m_{\pi})$.
This is done by examining the amplitude 
$\Lambda_{\mu\alpha\beta}(q,q')$ of \Eq{ArhorhoAmp} in the limit that the
photon momentum $Q \rightarrow 0$. Then, with the identity in
\Eq{WardId}, one can show that
\begin{eqnarray}
\lefteqn{
\lim_{Q\rightarrow 0} O^{[E]}_{\mu\alpha\beta} 
\Lambda_{\mu\alpha\beta}(q,Q\-q) 
} \nn \\
&=& 
O^{[E]}_{\mu\alpha\beta} \, 2 q_{\mu} {\bf T}_{\alpha\beta}(q,\-m_{\rho}^2)
\left(
\frac{d\;}{dq^2} \Pi^{PP}(q^2,m_{\pi},m_{\pi}) 
\right)_{\!\!q^2=-m_{\rho}^2}
\!\!\!, \!\!
\nn \\
&=& \delta Z_{\pi\pi},
\nn 
\end{eqnarray}
where $\delta Z_{\pi\pi}$ is that part of the contribution to the
normalization of the $\rho$ meson that arises from the two-pion
intermediate state: 
\begin{equation}
\delta Z_{\pi\pi} = 
\bigg(
\frac{d\;}{dq^2} \Pi^{PP}(q^2,m_{\pi},m_{\pi})
\bigg)_{q^2 = -m_{\rho}^2}
\label{deltaZpipi}
.
\end{equation}
It follows immediately that 
$G_{E}^{\pi\pi}(0) = \delta Z_{\pi\pi}$, which is the justification of the 
choice of normalizations in \Eq{Ge}.

From \Eq{Ge}, one may calculate the $\rho$ meson EM charge radius in the
usual manner,
\begin{eqnarray}
\langle r^2 \rangle &=&  - 6 \; \frac{d}{dQ^2} G_{E}(Q^2) 
\\ 
&=& \frac{-6}{1 + G_{E}^{\pi\pi}(0)} 
    \bigg( \frac{d}{dQ^2} G_{E}^{\bar{q}q}(Q^2) 
   \+ \frac{d}{dQ^2} G_{E}^{\pi\pi}(Q^2) \bigg)
.\!\!
 \label{RadiusDef}
\end{eqnarray}
The charge radius that results from a numerical calculation of
\Eq{RadiusDef} using \Eq{Gpipik2} is given in \Sec{Sec:Results}.

\section{Approximate Reduction to Time-Ordered Contributions}
\label{Sec:Orders}
The approach to evaluating the meson-loop corrections to the
$\rho$-$\omega$ mass splitting used here is to evaluate the covariant
Feynman diagram of Fig.~\ref{Fig:Loop}, which includes all possible
time orderings of the vertices and intermediate meson propagators. 
This is in contrast with the work of
Ref.~\cite{GI}, which evaluates the mass splitting using time-ordered
perturbation theory (TOPT), keeping only that part of the Feynman
diagram corresponding to two mesons propagating forward in time (the
forward-forward time ordering). In TOPT there is a separate
contribution from the two-meson Z graph, where the vertices are
ordered in such a way that the initial process (for, say, the $\rho$
self energy due to two-pion loops) is the creation of 
$\rho\pi\pi$ from the vacuum, which then propagate forward in time with the
original $\rho$ meson until the original $\rho$ meson annihilates with the
two $\pi$ mesons.  This is the backward-backward time ordering. 
A calculation of the forward time ordering requires knowledge of the
$\rho\pi\pi$ vertex at and near the vicinity of the real decay
kinematics, where it is reasonably well known.  However, a calculation of
the backward time ordering requires knowledge of the vertex for the
vacuum to $\rho\pi\pi$ process, which is not well known.  
Fortunately, the vertex for $\rho\pi\pi$ creation and annihilation is
believed to be strongly suppressed, and the backward-backward diagram is
further suppressed by the energy denominators of the $\rho\rho\pi\pi$
intermediate state propagation. 
For these reasons, the Z-diagram contribution is excluded from the
calculation described in Ref.~\cite{GI}.

With the belief that the multiple-hadron creation and annihilation
vertices should suppress the backward-backward time-ordered diagrams, a
simple calculation is undertaken to assess the relative importance of
this diagram to the forward-forward time ordering in the present covariant
framework. 
One might argue that predictions obtained within this covariant framework
would be neither robust nor reliable if the forward-forward and
backward-backward contributions were both large, but of opposite sign and
canceled each other.  Such subtle cancellations would be very sensitive 
to details of the model which may not be constrained accurately enough
to be trusted.  If there were a cancellation between the
forward-forward and backward-backward diagrams in the present framework,
it could explain the discrepancy between the small vector-meson mass shift
obtained herein, and the larger mass shifts obtained in the study of
\Ref{GI}, which keeps only the forward-forward contributions.

To determine the relative importance of the various time-ordered
contributions, one starts with the expression in \Eq{PiPP}, and then
rewrites the $\rho$-meson self energy as a sum of four contributions
arising from the four possible time orderings of the two pion propagators. 
One of these terms is identified with the contribution to the self energy 
due to the propagation of two {\em forward-propagating}
(positive-energy) pions. 
The analytic structure and physical interpretation of this term, which is
analogous to that calculated in \Ref{GI}, and the three other terms is 
discussed in detail.
In \Sec{Sec:ResultsOrders}, a direct numerical evaluation of these terms will
show that for time-like values of the vector-meson four momentum $q$, the
dominant contribution to the self energy $\Pi^{PP}(q^2)$ comes from the 
term identified with the propagation of two forward-propagating pions.
Therefore, although the inclusion of {\em all} four time orderings is
necessary to maintain the Lorentz covariance of the present framework, the
additional time-orderings present in covariant expressions like \Eq{PiPP},
are {\em not} responsible for the significant discrepancy between the mass
shifts obtained herein (to be presented in \Sec{Sec:ResultsOrders}) and
those obtained by \Ref{GI}.   
Rather, in the present quantum field theoretic framework, one observes
that the term in vector-meson self energy that arises from two
forward-propagating pions is, in fact, the dominant contribution.

Before proceeding further, it is necessary to clarify some issues
concerning the use of quantum field theoretic models formulated in 
Euclidean space.      
In the present study, observables are obtained by analytically
continuing the arguments of the Schwinger functions (Euclidean Green
functions) into Minkowski space.  In principal, this is done by
first integrating over the {\em internal} momenta so that the Schwinger
function depends only on Lorentz-scalar products of the external momenta.
The Schwinger function can then be analytically continued in these scalar
products into Minkowski space. 
For example, once the integration over the internal momenta $k$ is 
carried out in \Eq{PiPP}, the resulting Schwinger function $\Pi^{PP}(q^2)$
depends only on the square of the vector-meson four momentum $q^2$.
The Schwinger function is then analytically continued into Minkowski
space by letting $q^2 \rightarrow - m_{\rho}^2$.\footnote{
Additional details concerning the formulation of quantum field theory in
Euclidean space and the analytic properties of Schwinger functions may be
found in \Ref{Wightman}.}
In the present study, the use of model forms for the elementary
Schwinger functions (e.g., quark propagators, Bethe-Salpeter
amplitudes and meson transition vertices) allows the analytic continuation
of external momenta to be performed in a straight-forward manner by 
letting {external} momenta become complex before the {internal} momenta
are integrated over.  

An important ramification of such a framework is that a direct comparison
between the elementary Schwinger functions employed herein and those
employed in other models formulated in Minkowski space is impossible.
One reason for this is that the model form of the quark propagators
$S_f(k)$, given by \Eqs{QuarkProp}, are parametrized in terms of {\em
entire} functions of $k$.    
This parametrization provides the model with quark confinement by
ensuring that the quark propagator $S_f(k)$ is free of singularities in
the finite, complex-$k$ plane.  Hence, the quark propagator has no Lehman
representation and no quark-production thresholds can arise from the
calculation of observables.  
Of course, the use of such quark propagators also forbids the possibility
of performing a Wick rotation in the loop-integration variable $k$. 
Therefore, the elementary Schwinger functions employed herein are
permanently embedded in Euclidean space.  Only final quantities, such
as $\Pi^{PP}(q^2)$, can be analytically continued into Minkowski space. 

Another feature of the framework that makes a Wick rotation 
impossible to perform is the rich analytic structure of the
meson transition vertices, e.g., $\Lambda^{ijk}_{\mu}(p,p')$.
A Wick rotation of these transition vertices, although possible in
principal, is made difficult by the presence of singularities which arise
from the many-particle nature of the field theory.  
In general, these singularities may cross the integration path requiring
that exceptional care be taken when performing a Wick rotation.
In practice, these transition vertices are calculated and parametrized 
only on the small domain of the complex-momentum plane for which they are
needed in pion-loop integrations; their behavior outside this domain is
unknown.  
Such difficulties are not present in a quantum mechanical formulation that
truncates the Fock space in such a way as to include only states with a
small number of particles.   The analytic properties of transition
vertices obtained in such models are more easily analyzed. 

Nevertheless, it is possible to make a simple connection between this
field theoretic approach and TOPT.
The starting point is the calculation of the $\rho$-meson self energy due
to the two-pion loop from \Eq{PiPP}.
As in \Ref{GI}, the calculation is carried out in the rest frame of the
$\rho$ meson.  In this frame, $q_{\mu} = (\vec{0},im_{\rho})$ and the
$\pi$-meson propagators, which depend on the momenta  
$k_{\pm} = k \pm \sfrac{1}{2} q$, can be written in 
the following form 
\begin{eqnarray}
\Delta(k_{\pm},m_{\pi}) &=& 
	\frac{-1}{
	(ik_4 \mp \frac{1}{2}m_{\rho})^2 
	- (\omega(\vec{k}^2) - i \epsilon)^2}
\label{PiProp1},
\end{eqnarray}
where $\omega(\vec{k}^2) = \sqrt{\vec{k}^2 + m_{\pi}^2}$. 
The denominator of \Eq{PiProp1} is the difference of two squares. 
It can be factorized and rewritten as a difference of two poles in the
complex-$k_4$ plane, 
\begin{eqnarray}
\Delta(k^2_{\pm},m_{\pi}) =  \frac{1}{2 \omega(\vec{k}^2)}
  &\bigg[& 
  \frac{1}{ik_4 \mp \frac{1}{2}m_{\rho}	+ \omega(\vec{k}^2) - i \epsilon}
\nonumber \\
& & - 
  \frac{1}{ik_4 \mp \frac{1}{2}m_{\rho} - \omega(\vec{k}^2) + i \epsilon}
  \bigg]
\label{TOProp}.
\end{eqnarray}
In a field theory based in Minkowski space, the two terms in \Eq{TOProp} 
would be identified with forward- and backward-propagating pions.
A similar identification is possible for the two-pion system in our
approach by substituting \Eq{TOProp} into \Eq{PiPP}. 
One then obtains an expression for the vector meson self energy in which 
four poles in the complex-$k_4$ plane are explicit.  Writing the self
energy as a sum of four terms: 
$\Pi^{PP}(q^2)=\Pi^{[++]}(q^2)+\Pi^{[--]}(q^2)
+\Pi^{[+-]}(q^2)+\Pi^{[-+]}(q^2)$,
where, for $a, b = +,-$, one has   
\begin{eqnarray}
\Pi^{[ab]}(-m_{\rho}^2)
= 
-\frac{4}{3} \int\! \frac{d^4k}{(2\pi)^4}
\frac{{\cal F}^{PP}(k,q) }{4\omega^2(\vec{k}^2)}
\; A^{[ab]}(k) 
\label{TOPiPP},
\end{eqnarray}
and 
\begin{eqnarray}
A^{[++]}(k) &=&
\left(\frac{1}{ik_4-\frac{1}{2}m_{\rho}+\omega(\vec{k}^2)-i\epsilon}\right)
\nonumber \\ &&\times
\left(\frac{-1}{ik_4+\frac{1}{2}m_{\rho}-\omega(\vec{k}^2)+i\epsilon}
	\right) 
, \label{A++} \\ 
A^{[--]}(k) &=&  
\left(\frac{-1}{ik_4-\frac{1}{2}m_{\rho}-\omega(\vec{k}^2)+i\epsilon}\right)  
\nonumber \\ && \times 
\left(\frac{1}{ik_4+\frac{1}{2}m_{\rho}+\omega(\vec{k}^2)-i\epsilon}\right)
,  \\ 
A^{[+-]}(k) &=&  
\left(\frac{1}{ik_4-\frac{1}{2}m_{\rho}+\omega(\vec{k}^2)-i\epsilon}\right)
\nonumber \\ && \times 
\left(\frac{1}{ik_4+\frac{1}{2}m_{\rho}+\omega(\vec{k}^2)-i\epsilon}\right)
,  \\ 
A^{[-+]}(k) &=&  
\left(\frac{1}{ik_4-\frac{1}{2}m_{\rho}-\omega(\vec{k}^2)+i\epsilon}\right)
\nonumber \\ && \times 
\left(\frac{1}{ik_4+\frac{1}{2}m_{\rho}-\omega(\vec{k}^2)+i\epsilon}
	\right) .
\end{eqnarray}
These are identified as contributions to the $\rho$-meson self energy due
to forward-forward ($++$), backward-backward
($--$), forward-backward ($+ -$) and backward-forward ($-+$) propagation
of the two pions, respectively.
This nomenclature specifies the four possible time orderings of the two
pions in the pion loops.   It is justified from an analysis of the
motion of the poles in the complex-$k_4$ plane as the three-momentum
$\vec{k}$ is varied.  

\begin{figure}[ht]
 \begin{center}
 \epsfig{figure=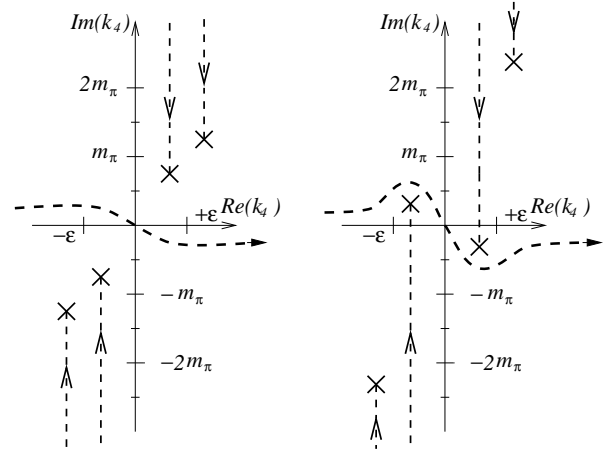,width=6.0cm,angle=-90}
 \end{center}
\caption{
The positions in the complex-$k_4$ plane of the poles in 
\Eq{TOPiPP}.
Dashed lines with arrows indicate the motion of these poles as the
relative three-momentum $|\vec{k}|$ is {\em decreased} from infinity to
zero. 
The positions of the poles for $|\vec{k}| = 0$ are denoted by ``X.''
The diagram on the left depicts the fictional situation for which 
$m_{\rho} \approx m_{\pi}/2 < 2 m_{\pi}$ and the $\rho \rightarrow \pi\pi$
decay is forbidden by energy conservation.
The diagram on the right depicts the situation for which $m_{\rho} > 2
m_{\pi}$ and the $\rho \rightarrow \pi \pi$ decay is allowed.
For a certain values of $|\vec{k}|$, two of the pion poles 
may pinch the $k_4$-contour integration as $\epsilon \rightarrow 0$.  
This signals the threshold of the $\rho \rightarrow \pi\pi$ decay and
results in a cut in the self energy of the $\rho$ meson. These two
poles correspond to the forward-forward time ordering in \Eq{A++}.
}
\label{Fig:Poles}
\end{figure}

These integrands in \Eq{TOPiPP} exhibit poles for particular values of the
fourth component of the loop-integration momentum $k_4$.
The positions of these poles depend on the magnitude of the three-momentum 
$\vec{k}$ as well as the mass of the vector meson $m_{\rho}$.   
In \Fig{Fig:Poles}, the positions of these poles, along with their motions
as $|\vec{k}|$ is varied, are depicted for two possible scenarios.
The left diagram corresponds to a scenario (not realized in nature)
for which the $\rho$-meson mass $m_{\rho} \approx m_{\pi}/2 < 2 m_{\pi}$.  
In this situation, the center-of-momentum energy of the $\rho$ meson is
below the two-pion decay threshold; hence, the $\rho$ meson is stable with
respect to the strong decay into two pions.
The diagram on the right depicts the situation (realized in nature) for
which the mass of the $\rho$ meson $m_{\rho} > 2 m_{\pi}$ and 
the strong, $\rho \rightarrow \pi \pi$ decay is allowed. 
In both diagrams, the motions of these poles as $|\vec{k}|$ is 
varied are denoted by dashed lines.  Arrows depict the trajectory of the
poles as $|\vec{k}|$ is {\em decreased} from infinity to zero.
The positions of the poles for $|\vec{k}| = 0$ are denoted by ``X.''

Examination of the left diagram in \Fig{Fig:Poles} depicts the fictional
situation where the $\rho$-meson mass is below the $\rho\rightarrow\pi\pi$
threshold energy.  Hence, the integrations over $k_4$ in \Eq{TOPiPP} never
encounter poles for any value of $|\vec{k}|$.  
Thus, in this scenario, the $\rho$-meson self energy $\Pi^{PP}(q^2)$ is
purely real and the $\rho$ meson is stable with respect to strong
interactions. 
In the right diagram of \Fig{Fig:Poles} for which $m_{\rho} > 2 m_{\pi}$, 
one observes a very different situation.  
Here, as the relative momentum $|\vec{k}|$ is decreased from $-\infty$, two
of the poles pass each other, thereby causing the contour integration
path for $k_4$ to become ``pinched'' in the limit that 
$\epsilon \rightarrow 0$. 
Such a pinching of the $k_4$-integration contour results in a branch cut
in $\Pi^{PP}(q^2)$ and hence a non-zero imaginary part of the vector-meson
self energy. 
This signals the opening of the two-pion decay channel and provides the 
$\rho$ meson with a finite two-pion decay width.
The value of $|\vec{k}|$ for which this pinch first develops is obtained
from the energy-conservation condition: $\omega(\vec{k}^2)=m_{\rho}/2$. 
The fact that this pinch singularity in $\Pi^{PP}(q^2)$ can only arise 
in the calculation of $\Pi_{++}(q^2)$ is the justification for this term
being identified with the propagation of two forward-propagating pions. 
Clearly, only positive-energy pions may appear in asymptotically free
states and in so doing contribute to the imaginary part of the
vector-meson  self energy.  

The term $\Pi^{--}(q^2)$ is identified as arising from two-backward
propagating pions by noting that had one started with the convention that
energies were negative rather than positive quantities, the replacement 
of the $\rho$-meson center-of-momentum energy $m_{\rho}$ with $-m_{\rho}$,
would result in the pinch singularity appearing in $\Pi^{--}(q^2)$
rather than in $\Pi^{++}(q^2)$.  
That is, within a framework that employed such a convention, the
asymptotically free states would necessarily contain negative-energy 
pions.  
A corollary that follows from the above argument is that in the limit that
$q^2 \rightarrow 0$, no energy flows into the two-pion-loop diagram and
there is no difference between the forward-forward and backward-backward
time orderings; that is, for a massless vector meson these time orderings
contribute equally to its self energy.  (This is indeed what is observed
numerically, as discussed in \Sec{Sec:ResultsOrders}.) 

It is important to note that the contribution to the vector-meson self
energy due to negative energy pions does indeed appear in quantum
mechanics, even though there are no negative energy states in the Fock
space.  By considering the time-ordered Feynman diagram associated
with $\Pi^{--}(q^2)$, written so that all particles move in the
positive-time direction, one observes that the four-particle Fock
state $|\rho\rho\pi\pi\rangle$ has a non-zero overlap with the
intermediate particles depicted in the diagram.  This state
contributes only to the real part of the vector-meson self energy
because energy conservation forbids the decay $\rho \rightarrow
\rho\rho\pi\pi$.

In a quantum mechanical framework, such as that employed in \Ref{GI}, 
Cauchy's Theorem entails that the mixed, forward-backward ($+-$) and
backward-forward ($-+$) contributions, are identically zero.  
However, in the present framework the meson-transition
vertex in \Eq{fVPPForm} and quark propagator in \Eq{QuarkProp} have 
essential singularities at infinity.  Therefore, one cannot close the
contour the integration at infinity and use Cauchy's theorem to show that
these terms are zero.  
Consequently, there is no such constraint on the value of these terms in
the present approach, and they will generally be different from zero.
A discussion of the physical interpretation of these terms and
the role they play in determining the unitarity of the theory is left for 
future investigations.
It is sufficient for the present application to note that in
\Sec{Sec:ResultsOrders}, a direct numerical evaluation of these terms
shows them to provide a negligible contribution to the vector-meson
self energy.  Hence, they have no impact on the results presented 
herein.    

\section{Meson transition amplitudes}
\label{Sec:Model}
In the previous sections, expressions for the vector-meson self energy
and EM form factors were given in terms of meson propagators and meson
transition amplitudes.  In the following section, these transition
amplitudes are given in terms of nonperturbative quark-loop
integrations that depend on model Bethe-Salpeter amplitudes and
nonperturbatively-dressed quark propagators developed in studies of
hadron observables based on the Schwinger-Dyson equations of QCD.
A recent review of such studies is provided by \Ref{Roberts}.

The imaginary part of the vector-meson self energy is determined by
meson transition amplitudes for which all external meson momenta are
on-mass-shell.  It follows that an experimental determination of decay
widths for the vector meson provides only the {\em on-mass-shell}
values of the meson transition amplitudes.  
In the following, the values of the transition amplitudes when all mesons
are on-mass-shell are referred to as {\em coupling constants}, e.g.,
$g_{\rho\pi\pi}$. 

A determination of the real parts of the vector meson self energy
requires calculating the principal part of the integrals in \Eqs{PiPP}
and (\ref{PiVP}), which sample the $k^2$-integration domain for which
the intermediate mesons are off-mass-shell.  Thus, a calculation of
the real parts of the self energy and EM form factors requires
knowledge of
the meson transition form factors for off-mass-shell values of the
intermediate meson momenta.  As the real part of the self energy is
not directly observable, experiment cannot determine the behavior of
the off-mass-shell-meson transition form factors.  Nonetheless,
experimental observations can provide some guidance in helping to
constrain the real parts of self energies and EM form factors.  This can
be done by looking for observables which are proportional to the {\em
difference} of two self energies.  For example, from \Eqs{PiSum:Rho}
and (\ref{PiSum:Omega}), it is apparent that the $\rho$--$\omega$ mass
splitting provides a measure of the difference of the real parts of
the $\rho$ and $\omega$ self energies.
In \Sec{Sec:Model}, the calculation of off-mass-shell meson
transition form factors is discussed.  

%
%
The VVP and VPP vector-meson transition amplitudes are calculated
within a generalized impulse approximation in which the mesons couple
to each other by means of a nonperturbatively-dressed quark loop.  In
this generalized impulse approximation, the quark propagators and
meson Bethe-Salpeter amplitudes are nonperturbatively dressed, but explicit
three-body interactions are neglected.

In the following, quark-loop integrations are performed in Euclidean
space (where $q^2 > 0$), and final results are analytically continued
to Minkowski space by letting the external (timelike) momenta $q^2
\rightarrow - m^2_{V}$.  The Dirac matrices $\gamma_{\mu}$ employed
herein are Hermitian, $\gamma_{\mu}^{\dag} = \gamma_{\mu}$, and
satisfy the anti-commutation relation $\{ \gamma_{\mu},\gamma_{\nu} \}
= 2 \delta_{\mu \nu}$.

The amplitude describing the coupling of a vector meson of momentum
$q = -p_1-p_2$ and two pseudoscalar mesons with momentum $p_1$ 
and $p_2$ is given in impulse approximation by\footnote
{The sense of all momenta is such that positive momenta flow {\em into}
vertices.},  
\begin{eqnarray}
\lefteqn{
\Lambda^{ijk}_{\mu}(p_1,p_2) 
= {\rm tr}_{CFD}\int \! \frac{{\rm d}^4k}{(2\pi)^4} 
{\bf S}(k_{++}) 
\lambda_{i} V_{\mu}(k\+\sfrac{1}{2}p_2;q) 
}
\nn \\
& & \times {\bf S}(k_{-+})
      \lambda_{j} \Gamma_{P}(k\-\sfrac{1}{2}q;p_1) 
{\bf S}(k_{--})  \lambda_{k} \Gamma_{P}(k;p_2)
,
\label{LamVPP:Quark}
\end{eqnarray}
where $k_{\alpha\beta} = k + \sfrac{\alpha}{2} q + \sfrac{\beta}{2} p_2$, 
$\lambda_i$ are linear combinations of the Gell-Mann
SU(3)-flavor matrices associated with the mesons involved, the trace
is over color, flavor and Dirac indices, and ${\bf S}(k) = {\rm
diag}(S_u(k),S_d(k),S_s(k))$ is the quark propagator for $u$, $d$ and
$s$ quarks. It is a 3$\times$3 matrix in the quark-flavor space.  In
this study, the effects of SU(3)-flavor breaking are neglected, so the
quark propagator is diagonal in flavor indices and $S_u(k) = S_d(k) =
S_s(k)$.  Here $V_{\mu}(k,k')$ is the vector-meson Bethe-Salpeter
amplitude for the $\rho$ or $\omega$ mesons, and $\Gamma_{P}(k;p_1)$
is the Bethe-Salpeter amplitude for the $P$ = $\pi$ or $K$ mesons.
The flavor structure of the meson Bethe-Salpeter amplitudes is given
explicitly in \Eq{LamVPP:Quark} in terms of the 3$\times$3 Gell-Mann
matrices $\lambda_i$ in the quark-flavor space.
The meson transition amplitude $\Lambda_{\mu}^{ijk}(p_1,p_2)$ is depicted
in \Fig{Fig:Loop} as a triangle diagram. 

The transition form factor defined by \Eq{VPPvertex} is obtained from
\Eq{LamVPP:Quark} by choosing values of the flavor indices $i$, $j$
and $k$ appropriate for the particular process $V\rightarrow PP$ under
consideration, and then contracting the amplitude with an appropriate
vector.  In the case of $\rho^{0} \rightarrow \pi^{+} \pi^{-}$, one
obtains
\begin{equation}
g_{\rho\pi\pi} f^{VPP}(p_1,p_2) 
= \frac{ (p_1 + p_2)_{\mu}}{m_V^2 - 4 m_P^2}
\Lambda^{3-+}_{\mu}(p_1,p_2)
\label{fVPP_Proj}
,
\end{equation}
where the flavor indices refer to the following linear
combinations of Gell-Mann matrices $\lambda_{\pm} =
\sfrac{1}{\sqrt{2}}( \lambda_1 \pm i \lambda_2)$, which are associated
with an isovector meson of good charge.  The resulting form factor
$f^{VPP}(p_1,p_2)$ is the essential dynamical element
necessary to calculate the real part of the two-pseudoscalar
intermediate state contribution to the vector-meson self energy
$\Pi^{PP}(q^2)$ from \Eq{PiPP}.

Transition amplitudes for processes involving strange
mesons, such as $\rho \rightarrow K \bar{K}$, are obtained by assuming
SU(3)-flavor symmetry is exact, so that $S_s(k) = S_u(k)$ and
$\Gamma_{K}(k;p_1) = \Gamma_{\pi}(k;p_1)$.  From considerations of
SU(3)-flavor symmetry, one can show that the left-hand-side of
\Eq{fVPP_Proj} gains an additional factor of $1/\sqrt{2}$ for the
process $\rho \rightarrow K \bar{K}$.  In particular, for $\rho^+
\rightarrow K^+ \bar{K}^0$, the analogous result for \Eq{fVPP_Proj}
would be
\begin{equation}
\frac{ g_{\rho\pi\pi}}{\sqrt{2}}f^{VVP}(p_1,p_2) 
= \frac{ (p_1 + p_2)_{\mu}}{m_V^2 - 4 m_P^2} 
\Lambda^{3 K^{-} {K}^0}_{\mu}(p_1,p_2)
\label{fVPP_Proj2}
.
\end{equation}
Here, the flavor indices denote the linear combinations of Gell-Mann
matrices, 
$\lambda_{K^{-}} \equiv \sfrac{1}{\sqrt{2}} ( \lambda_4 - i \lambda_5)$
and 
$\lambda_{{K}^{0}} \equiv \sfrac{1}{\sqrt{2}} ( \lambda_6 + i \lambda_7)$.
The additional factor of $1/\sqrt{2}$ that appears in \Eq{fVPP_Proj2}
relative to that in \Eq{fVPP_Proj}, is the reason for the factor of $1/2$
that appears as the coefficient of $\Pi^{PP}(q^2,m_K,m_K)$ in 
\Eqs{PiSum:Rho} and (\ref{PiSum:Omega}).

Similarly, the coupling of two vector mesons of momenta $-(p_1+p_2)$
and $p_1$ to a pseudoscalar meson of momentum $p_2$ is given in
the impulse approximation by
\begin{eqnarray}
\lefteqn{
\Lambda^{ijk}_{\mu\nu}(p_1,p_2) 
= 
{\rm tr}_{CFD}\int \! \frac{{\rm d}^4k}{(2\pi)^4} 
{\bf S}(k_{++})  
\lambda_{i} V_{\mu}(k\+\sfrac{1}{2}p_2;q) 
} 
\nn \\
& &
\times {\bf S}(k_{-+})   \lambda_{j} V_{\nu}(k\-\sfrac{1}{2}q;p_1) 
{\bf S}(k_{--})  \lambda_{k} \Gamma_{P}(k;p_2)
.
\label{LamVVP:Quark}
\end{eqnarray}
The elements appearing in \Eq{LamVVP:Quark} are the same as in
\Eq{LamVPP:Quark}. No new elements are necessary to obtain the $VVP$
transition amplitude.  The form factor $f^{VVP}(p_1,p_2)$ is obtained from 
\Eq{LamVVP:Quark} using \Eq{VVPvertex} and is the essential
dynamical element necessary to calculate the real part of the intermediate
vector-pseudoscalar contribution to the vector-meson self energy 
$\Pi^{VP}(q^2,m_V,m_P)$ from \Eq{PiVP}.

Here too, the assumption of exact-SU(3)-flavor invariance provides a
simple relation between the processes 
$\rho \rightarrow \omega \pi$, 
$\rho \rightarrow \rho \eta$,
$\rho \rightarrow K^* K$, 
$\omega \rightarrow \rho \pi$, 
$\omega \rightarrow \omega \eta$,
and $\omega \rightarrow K^* K$.
The flavor factors that relate these processes lead to the
coefficients appearing in front of the vector-pseudoscalar contributions 
in \Eqs{PiSum:Rho} and (\ref{PiSum:Omega}).

Evaluation of the quark-loop integrations in \Eqs{LamVPP:Quark} and
(\ref{LamVVP:Quark}) requires knowledge of the $u$-, $d$- and
$s$-quark propagators as well as the vector-meson and pseudoscalar-meson 
Bethe-Salpeter amplitudes.  These are taken from phenomenological studies
of the electromagnetic form factors and strong and weak decays of pseudoscalar
and vector mesons \cite{Hawes,Burden}. 

The form of the model, dressed-quark propagators are based on several
numerical studies of the quark Schwinger-Dyson equation using a realistic
model-gluon propagator \cite{Frank,Maris}.
In a general covariant gauge, the propagator for a quark of flavor $f$ is
written as $S_f(k) = - i \gamma \cdot k \sigma_V^f(k^2) +
\sigma_S^f(k^2)$.
It is well described by the following parametrization:
\begin{eqnarray}
  \bar{\sigma}^f_S(x) & = & 
    \xi[b^f_1 x] \; \xi[b^f_3 x]
    \left(b^f_{0} + b^f_2 \xi[ \Lambda x]  \right)
      \nonumber\\    
   & &  
    + 2 \bar{m}_f \xi\big[ 2(x+\bar{m}_f^2) \big] 
	+ C_{f} e^{-2x}\:, \nonumber\\
  \bar{\sigma}^f_V(x) & = & 
  \frac{ 2(x+\bar{m}_f^2) - 1 + e^{-2(x+\bar{m}_f^2)} }
       { 2(x+\bar{m}_f^2)^2 }\:, 
	\label{QuarkProp}
\end{eqnarray}
where $\xi[x] = (1 - e^{-x})/x$, 
$x = k^2/\lambda^2$, $\bar{\sigma}^f_{S} = \lambda \sigma^f_S$,
$\bar{\sigma}^f_V = \lambda^2 \sigma^f_V$, 
$\bar{m}_f = m_f / \lambda$, 
$\lambda =$ 0.566~GeV,
$\Lambda = 10^{-4}$, 
and where the $b_i^f$, $C_{f}$ and $m_f$ are parameters.  Since
$\xi[x]$ is an entire function, this dressed-quark propagator has {\em
no} Lehmann representation.  The lack of a Lehmann representation for
$S_f(k)$ is sufficient to ensure that the quarks are confined, since
no quark-production thresholds can appear in the calculation of
observables.  The quark propagator $S_f(k)$ also reduces to a
bare-fermion propagator when its momentum is large and spacelike, in
accordance with the asymptotic behavior expected from perturbative QCD
(up to logarithmic corrections).  The parameters for the $u$-, $d$-
and $s$-quark propagators were determined in \Ref{Burden} by
performing a $\chi^2$ fit to a range of $\pi$- and $K$-meson
observables.  The parameters for the $s$-quark propagator employed
herein were re-fit in \Ref{Hawes}, also from a $\chi^2$ fit to
$K$-meson observables, but with the additional constraint that the
resulting $s$-quark propagator more closely resemble that obtained by
the numerical study of \Ref{Maris}.  The resulting parameters are
given in \Tab{Tab:QuarkParam}.

\begin{table}
\begin{center}
\caption{Confined-quark propagator parameters for $u$, $d$, and $s$
quarks from \Ref{Hawes}.  An entry of ``same'' for a meson indicates
that the value of the parameter is the same as for the quark flavor
with which it is grouped.}
\begin{tabular}{l|lllllc}
$f$/meson& $b^f_0$ & $b^f_1$ & $b^f_2 $ & $b^f_3$ & $C_{f}$ & $m_f$
[MeV]\\ \hline
$u,d$ &   0.131 &   2.900 &   0.603  & 0.185 &  $\;\;$0  &   5.1 \\
$\pi$ &   same  &   same  &   same   & same  &  0.121    & $0\;\;\;$ \\
$\rho$, $\omega$&   
          0.044 &   0.580 &   same   & 0.462 &  $\;\;$0  & $0\;\;\;$ \\ \hline
$ s $ &   0.105 &   2.900 &   0.540  & 0.185 &  $\;\;$0  & 130.0 \\
$ K $ &   0.322 &   same  &   same   & same  &  $\;\;$0  & $0\;\;\;$ \\
$K^*$ &   0.107 &   0.870 &   same   & 0.092 &  $\;\;$0  & $0\;\;\;$ \\
\end{tabular}
\label{Tab:QuarkParam}
\end{center}
\end{table}

In the chiral limit $m_f \rightarrow 0$, Goldstone's theorem allows one
to obtain the pseudoscalar-meson Bethe-Salpeter 
amplitude $\Gamma_{P}(k ; P)$ directly
from  the dressed-quark propagator $S_f(k)$ \cite{Maris}.
For the calculation of infrared transition amplitudes, like those studied
herein, it is sufficient to keep only the part of the pseudoscalar
Bethe-Salpeter amplitude that is proportional to $\gamma_5$.  In this
case, the amplitude is given by
\begin{equation}
  \Gamma_{P}(k ; P)
  = \gamma_5 \:
   \frac{ B_P(k^2) }{N_{P}} 
\:,
  \label{PiBSAmp}
\end{equation}
where $B_P(k^2)$ has the same form as $B_f(k^2)$, one of the Lorentz
invariants appearing in the quark inverse propagator $S_f^{-1}(k) = i
\gamma \cdot k A_f(k^2) + B_f(k^2)$, but with the parameters given in
\Tab{Tab:QuarkParam} for $P = \pi$ or $K$.  The $\pi$- and $K$-meson
Bethe-Salpeter amplitudes are obtained from Eqs.~(\ref{QuarkProp})
and(\ref{PiBSAmp}) using the parameters from \Tab{Tab:QuarkParam}.

In \Eq{PiBSAmp}, $N_P$ is the on-mass-shell normalization for the $P=$
$\pi$ or $K$ meson, and is determined by the normalization condition: 
\begin{eqnarray}
1 &=& \frac{ N_c p_{\mu}}{m_{P}} 
{\rm tr}_{D} \int\! \frac{{\rm d}^4 k}{(2\pi)^4} 
\nn \\
& & 
\frac{\rm d\;}{{\rm d}p_{\mu}} S(k\+\sfrac{1}{2}p)
\Gamma_{P}(k;p)
S(k\-\sfrac{1}{2}p) \bar{\Gamma}_{P}(k;p) 
\nn\ \\
& &+ 
S(k\+\sfrac{1}{2}p) {\Gamma}_{P}(k;p) 
\frac{\rm d\;}{{\rm d}p_{\mu}} S(k\-\sfrac{1}{2}p)
\bar{\Gamma}_{P}(k;p)
,\label{fK}
\end{eqnarray}
where ${\rm tr}_{D}$ denotes a trace over Dirac indices and $N_c = 3$.  
In \Ref{Maris} it
was shown that by maintaining only the Dirac amplitude proportional to
$\gamma_5$ in the pion Bethe-Salpeter amplitude, one cannot reproduce the
ultraviolet behavior of the pion electromagnetic form factor. To do so
requires the inclusion of the axial-vector $\gamma_5 \gamma \cdot p$
components into the $\pi$-meson Bethe-Salpeter amplitude.  Nonetheless, for the
calculation of observables which involve small momentum transfers,
neglecting Dirac moments other than $\gamma_5$ and normalizing the
Bethe-Salpeter amplitude according to \Eq{fK} is usually sufficient.

The vector meson Bethe-Salpeter amplitudes employed herein are taken from
\Ref{Hawes}:   
\begin{equation}
  V_{\mu}(k; p)  = \left( \gamma_{\mu} + \frac{p_{\mu} \gamma \cdot
    p}{m_V^2} \right) 
	 \frac{ B_V(k^2) }{N_{V}} 
\:,
 \label{VectorBSAmp}
\end{equation}
where $p$ is the momentum of the vector meson with mass $m_V$,
$k$ is the relative momenta of the quark and antiquark, 
$N_V$ is the Bethe-Salpeter normalization, obtained from an obvious
generalization of \Eq{fK}, and $B_V(k^2)$ has the same form as the
Lorentz-invariant 
function $B_f(k^2)$ found from the quark propagator of \Eq{QuarkProp}
with parameters given in \Tab{Tab:QuarkParam} with $V = \rho$, 
$\omega$ or $K^*$.

Again, all but the leading Dirac moment proportional to $\gamma_{\mu}$
in the vector meson Bethe-Salpeter amplitude have been neglected.  A
numerical study \cite{BQRTT} of the Bethe-Salpeter equation that employed a
separable model for the quark-antiquark scattering kernel found that
the magnitude of this moment is about ten times larger than the next
largest.  However, some observables may be more sensitive to
sub-leading Dirac moments in the Bethe-Salpeter amplitude.  Indeed, this
was the case observed in \Ref{Maris} for the asymptotic-$q^2$ behavior of
the $\pi$ meson electromagnetic form factor.  However, the simple form
provided by \Eq{VectorBSAmp} is sufficient to provide a reasonable
model for calculating the off-mass-shell transition form factors
$f^{VPP}(p_1,p_2)$ and $f^{VVP}(p_1,p_2)$.

With these elements having already been determined by previous
studies, the on- and off-mass-shell values of the transition
amplitudes $\Lambda^{VPP}_{\mu}(p_1,p_2)$ and $\Lambda^{VVP}_{\mu
\nu}(p_1,p_2)$ can be calculated with no adjustable parameters.  The
value of a transition amplitude for on-mass-shell meson momenta is
referred to as a ``coupling constant''.  If the invariant mass of the
final state $\sqrt{s}$ is less than the mass of the initial state, the
decay is allowed to proceed and the coupling constant is
experimentally observable.  If energy conservation prevents all
external mesons from being simultaneously on-mass-shell, one can still
define the ``coupling constant'' as the value of the transition
amplitude at this point.  However, in this case, there is no decay
from which it can be measured experimentally.

In \Ref{Hawes}, the coupling constants $g_{\rho \pi \pi}$ and $g_{K^*
K\pi}$ were obtained using the model quark propagators from
\Eqs{QuarkProp} and Bethe-Salpeter amplitudes from \Eqs{PiBSAmp} and
(\ref{VectorBSAmp}) with the parameters from \Tab{Tab:QuarkParam}.
The values for $g_{\rho \pi \pi}$ and $g_{K^* K\pi}$ obtained therein
were 8.52 and 9.66, respectively.  The experimentally observed values
for these coupling constants are $6.03\pm0.02$ and $6.40\pm0.04$,
respectively.  There are several possible reasons that this model
gives results for the decay widths of a vector meson into two
pseudoscalars that are larger than obtained by experiment.  The first
is that for this process in particular, higher Dirac moments
contribute significantly to the magnitude of the coupling constant
\cite{MAP2}.  The second is the possibility that final-state
interactions may also be important.  For these reasons, rather than
employ the value of $g_{\rho \pi \pi}$ obtained in \Ref{Hawes}, the
experimentally determined value of $g_{\rho \pi \pi} = 6.03$ is used.
The meson transition form factors $f^{VPP}(p_1,p_2)$ and
$f^{VVP}(p_1,p_2)$ are the important dynamical element necessary for
the present calculations.  The scales of these are rather insensitive to
details of the model Bethe-Salpeter amplitudes, so that the model of
\Ref{Hawes} is quite adequate for their determination.

The coupling constant $g_{\rho \omega \pi}$ = 10.81 is obtained from
the $\rho\omega\pi$ transition amplitude in \Eq{LamVVP:Quark}, when
all three mesons are on-mass-shell.  Of course, the lack of phase
space makes the decay $\omega \rightarrow \rho \pi$ impossible, so
that the vector-meson self energy contribution due to
vector-pseudoscalar intermediate meson states is real for a physical
vector meson; i.e., the value of $g_{\rho \omega \pi}$ cannot be
determined experimentally.  Throughout this article, the value
$g_{\rho \omega \pi}$ = 10.81 obtained from \Eqs{PiVP} with
(\ref{FVP}) is used \cite{MAP3}.  For comparison, the value of this
coupling constant can been estimated from the anomalous Wess-Zumino
term in an effective chiral Lagrangian with the assumption of a
universal $\rho$-meson coupling.  With the definition of the coupling
constant from \Eq{VVPvertex}, it is found to be $g_{\rho \omega \pi}
\approx 11.5$ \cite{UMeissner}, which is consistent with the results
of the present calculation. Similarly, a study of the decays of the
$\omega$-meson using vector-meson dominance finds the value 
$g_{\rho \omega \pi} \approx 9.7$~\Ref{Durso}.

%
The loop integrations of Eqs.~(\ref{PiPP}) and~(\ref{PiVP}) sample the
meson transition form factors $f^{VPP}(p_1,p_2)$ and $f^{VVP}(p_1,p_2)$ 
for {\em off-mass-shell} values of the intermediate-meson four momenta $p_1$
and $p_2$.  A calculation of these form factors from the quark-loop
integrations of \Eqs{LamVPP:Quark} and (\ref{LamVVP:Quark}) requires a
suitable definition for meson Bethe-Salpeter amplitudes when their total
momentum is off-mass-shell.  
Although Bethe-Salpeter amplitudes are uniquely determined from the
homogeneous Bethe-Salpeter equation for on-mass-shell values of the
total momentum, an extrapolation of the amplitude away from this
on-mass-shell value is {\em not} uniquely determined, and hence, is
model dependent.

Herein, the Bethe-Salpeter amplitudes as defined in \Eqs{PiBSAmp} and
(\ref{VectorBSAmp}) are used, even for values of the meson four
momenta that are off-mass-shell.  This is most similar to the
procedure employed in quark model calculations such as \Ref{GI}, where
a complete set of intermediate states is inserted between the
vector-meson states leading to transition amplitudes for on-shell
mesons.  The off-shell behavior of the intermediate mesons is then
determined solely by the behavior of the meson propagators.

\begin{figure}[t]
 \begin{center}
 \epsfig{figure=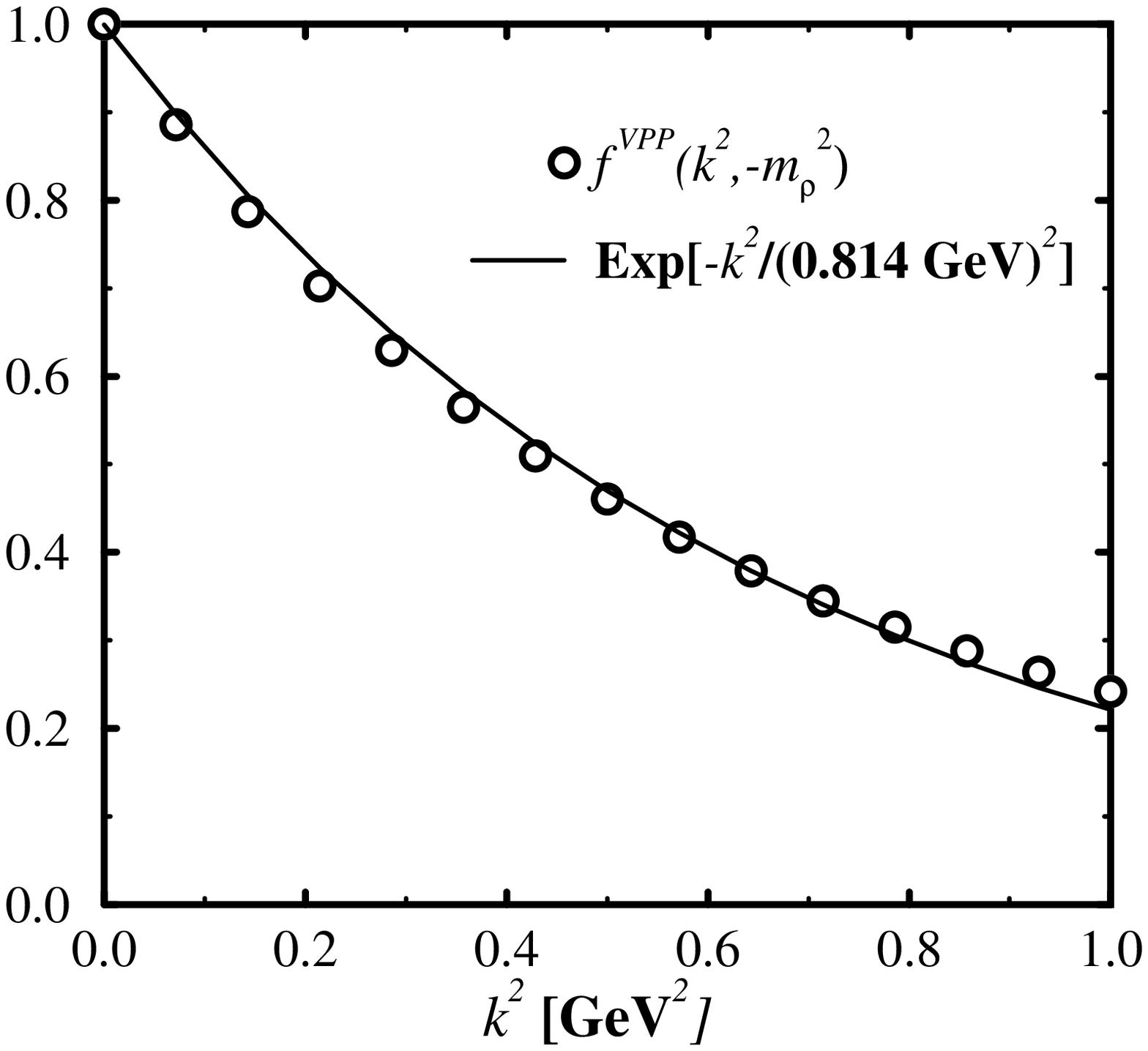,height=7.0cm,width=7.0cm}
 \end{center}
\caption{
The meson transition form factor $f^{VPP}(p_1,p_2)$ 
obtained from \Eqs{LamVPP:Quark} (circles) and the numerical fit (solid curve)
of \Eq{VPPFit} used in the calculations of vector meson self energies and
EM form factors.  
}
\label{Fig:fVPP}
\end{figure}

\begin{figure}[t]
 \begin{center}
 \epsfig{figure=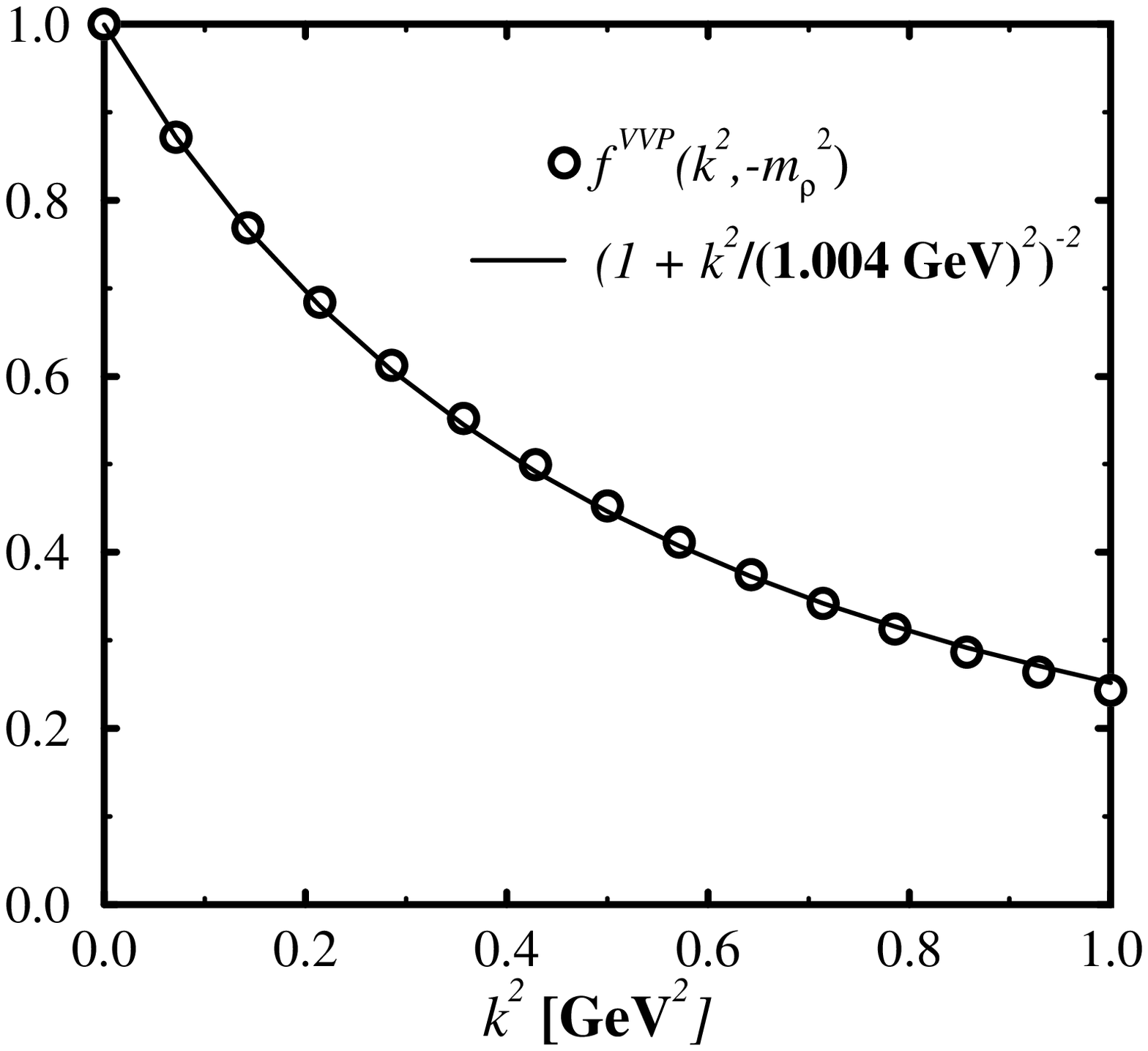,height=7.0cm,width=7.0cm}
 \end{center}
\caption{
The meson transition form factor $f^{VVP}$ 
obtained from \Eqs{LamVVP:Quark} (circles) and the numerical fit (solid curve)
of \Eq{VVPFit} used in the calculations of vector meson self energies.  
}
\label{Fig:fVVP}
\end{figure}

Rather than directly use the numerical values of the meson transition
form factors as calculated from \Eqs{LamVPP:Quark} and
(\ref{LamVVP:Quark}), it is found that the resulting form factors can
be parametrized in terms of the following functions:
\begin{eqnarray}
f^{VPP}(k\+\frac{q}{2},\-k\+\frac{q}{2}) &=& 
  \exp\left(-\frac{k^2 \+ m_{P}^2 \- \sfrac{1}{4}m_{V}^2}
{b_{VPP}^2(q^2)}\right) , 
\label{fVPPForm}  \\
f^{VVP}(k\+\frac{q}{2},\-k\+\frac{q}{2}) &=& 
   \left(
    \frac{1 \- m_{P}^2/b_{VPP}^2(q^2)}{1 \+ k^2/b_{VPP}^2(q^2)}
  \right)^2 
\label{fVVPForm},
\end{eqnarray}
where $p_1 = \sfrac{1}{2} q + k$ and $p_2 = \sfrac{1}{2} q - k$ in
\Eq{fVPPForm} and \Eq{fVVPForm}.  The parameters are chosen by fitting
these forms to the numerical results on the domain $k^2 \in$ (0,
1)~GeV$^2$ for a range of external vector-meson four momenta $q^2 \in$
($-1$, $+1$)~GeV$^2$.  The $q^2$-dependence of the resulting
parameters is well described in terms of the linear forms:
\begin{eqnarray}
b_{VPP}(q^2) &=& b_{VPP}^0 + b_{VPP}^1 \; q^2 ,
\nn \\
b_{VPP}^0 &=& 0.855 \;\;{\rm GeV} ,
\nn \\
b_{VPP}^1 &=& 0.0693 \;\;{\rm GeV}^{-1}, 
	\label{VPPFit} 
\end{eqnarray}
and
\begin{eqnarray}
b_{VVP}(q^2) &=& b_{VVP}^0  + b_{VVP}^1 \; q^2 ,
\nn \\
b_{VVP}^0 &=& 1.098 \;\;{\rm GeV} ,
\nn \\
b_{VVP}^1 &=& 0.159 \;\;{\rm GeV}^{-1}
\label{VVPFit}
.
\end{eqnarray}
The numerical results (denoted by circles) and the fits (solid curves)
are shown in Figs.~\ref{Fig:fVPP} and \ref{Fig:fVVP} for $q^2 =
-m_{\rho}^2$ for the form factors $f^{VPP}(p_1,p_2)$ and $f^{VVP}(p_1,p_2)$,
respectively.  The similarity of the two curves suggests that both
could have just as easily been fitted to either exponential or dipole
forms.  The choice of one form over the other is arbitrary, and does
not affect the results obtained herein.

\section{Results and discussion}
\label{Sec:Results}

\subsection{mass dependence}
\label{Sec:ResultsMassDep}
%
\begin{figure}[t]
\begin{center}
\epsfig{figure=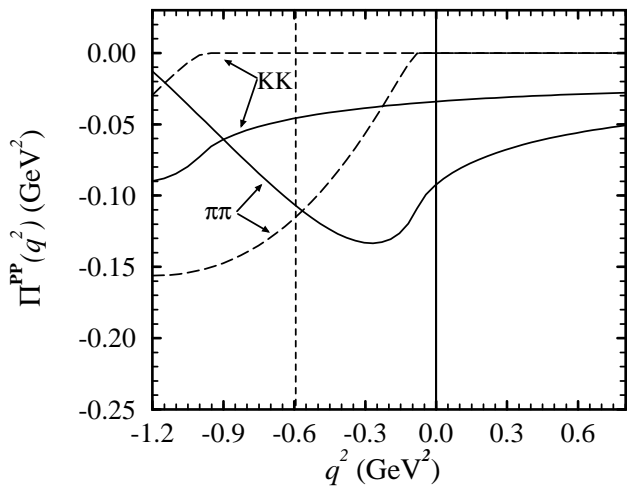,height=7.0cm}
\end{center}
\caption{
The real (solid curves) and imaginary (dashed curves) parts of
the vector-meson self energy due to intermediate $\pi\pi$ and $KK$ meson
states.
The vertical dashed line denotes the on-mass-shell point $q^2 = -
m_{\rho}^2$ for the $\rho$ meson 
and the vertical solid line denotes $q^2 = 0$. 
}
\label{Fig:VPP}
\end{figure}

\Fig{Fig:VPP} shows the real (solid curves) and imaginary
(dashed curves) parts of the contributions to the vector self energy
due to $\pi\pi$ and $KK$ loops as a function of the four-momentum squared
of the vector meson $q^2$. 
These curves are obtained by calculating $\Pi^{PP}(q^2,m_P,m_P)$ 
with $P=$ $\pi$ or $K$ in \Eqs{PiPP} and (\ref{FPP}) with the
assumption of exact SU(3)-flavor symmetry; i.e., 
$g_{\rho\pi\pi} = g_{\rho K K} =$ 6.03 and using the parametrization of
$f^{VPP}(p_1,p_2)$ given by \Eq{fVPPForm}.

For spacelike values of the vector-meson momenta $q^2 > 0$ (to the
right of the solid vertical line in \Fig{Fig:VPP}), the self energy
$\Pi^{PP}(q^2,m_P,m_P)$ is purely real and approaches zero
monotonically from below as $q^2 \rightarrow + \infty$.  The threshold
for the decay $\rho \rightarrow\pi\pi$ is at $q^2 = - 4 m_{\pi}^2$
(just left of the solid vertical line in \Fig{Fig:VPP}).  Here, the
imaginary part of the self energy $\Pi^{PP}(q^2,m_{\pi},m_{\pi})$
becomes non-zero.  This abrupt change of the imaginary part of the
self energy at the two-pion threshold leads to a discontinuity in the
{\em second derivative} of the real part of the self energy.  This
causes the real part of the self energy to rapidly turn over at 
$q^2 \approx$ $-0.3$~GeV$^2$ and start to approach zero as $q^2$ becomes
more negative (timelike).  The same behavior is observed in the
two-$K$-meson self energy $\Pi^{PP}(q^2,m_{K},m_{K})$, but much deeper
in the timelike region since the two-kaon decay threshold is at $q^2 =
-4 m^2_{K} \approx$ $-1$~GeV$^2$.

The discontinuity in the second derivative of 
${\rm Re}\Pi^{PP}(q^2,m_{P},m_{P})$ at $q^2 = -4 m_{P}^2$ is a general
feature of an amplitude near the threshold of a decay into a two-meson
intermediate state with $L=1$ relative angular momentum. 
In general, the dependence of the imaginary part of the self energy on the
center-of-momentum energy $\sqrt{s}$ near a decay threshold is
proportional to $\sqrt{s}^{(1 + 2 L)}$, where $L$ is the relative angular
momentum of the two outgoing decay products.  
The centrifugal barrier due to the relative angular momentum $L$ 
tends to soften the behavior of the imaginary part of
the amplitude at threshold.   
In the present case of $\rho\rightarrow\pi\pi$, the pions have 
relative $L=1$ angular momentum and hence the second derivative of 
${\rm Re}\Pi^{PP}(q^2,m_{P},m_{P})$ is discontinuous at $q^2 = -4 m_{P}^2$.
Had this two-body decay process proceeded in the $s$-channel with $L=0$,
the {\em first} derivative of the self energy would have been observed as
being discontinuous at threshold $q^2 = -4 m_{P}^2$.

\begin{figure}[t]
\begin{center}
\epsfig{figure=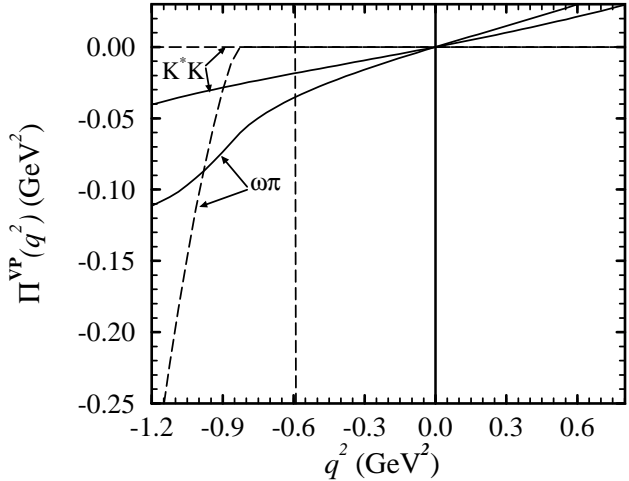,height=7.0cm}
\end{center}
\caption{
The real (solid curves) and imaginary (dashed curves) parts of
the vector-meson self energy due to intermediate $\omega\pi$ 
and $K^* K$ meson states.
The vertical dashed line denotes the on-mass-shell point $q^2 = -
m_{\rho}^2$ for the $\rho$ meson 
and the vertical solid line denotes $q^2 = 0$. 
}
\label{Fig:VVP}
\end{figure}
In \Fig{Fig:VVP} the real (solid curves) and imaginary
(dashed curves) parts of the vector-meson self energy due to
vector-pseudoscalar intermediate state $\Pi^{VP}(q^2,m_V,m_P)$ is shown as
a function of $q^2$.
These curves are obtained by calculating $\Pi^{VP}(q^2,m_V,m_P)$ 
with $P=\pi$ and $V=\rho$ or $P=K$ and $V=K^*$ in \Eqs{PiVP} and
(\ref{FVP}) with the assumption of exact SU(3)-flavor symmetry; i.e., 
$g_{\rho\omega\pi} = g_{\rho K^* K} =$ 10.81 and using the parametrization
of $f^{VVP}(p_1,p_2)$ given by \Eq{fVVPForm}.

Although not apparent from \Fig{Fig:VVP}, when the vector-meson four
momentum becomes large and spacelike, $q^2 \rightarrow +\infty$, the real
part of the self energy $\Pi^{VP}(q^2,m_V,m_P)$ approaches zero from above.
The fact that the decay threshold for $\rho\rightarrow\omega\pi$ is
observed at $q^2 = -(m_{\omega}+m_{\pi})^2 \approx$ -0.84~GeV$^2$, so
far above the on-mass-shell value of $q^2 = -m_{\rho}^2$ means that very
little structure for the real part of $\Pi^{VP}(q^2,m_{\omega},m_{\pi})$ 
is observed for the range of $q^2$ plotted in \Fig{Fig:VVP}.
The lack of structure in the real part of  $\Pi^{VP}(q^2,m_{K^*},m_{K})$
for the $K^*K$ channel is even more apparent, owing to the fact that the
threshold for $\rho \rightarrow K^* K$ is found at 
$q^2 = -(m_{K^*}+m_{K})^2 \approx$ -1.9~GeV$^2$ (not shown in
\Fig{Fig:VVP}). 

As discussed above for $\Pi^{PP}(q^2,m_P,m_P)$, the order in which the
discontinuity in derivatives of the real part of the self energy
appears is determined by the relative angular momentum $L$ of the
out-going decay products.  It is straight-forward to show that the
decay of a vector meson with $J^P = 1^-$ into a vector and
pseudoscalar meson with $J^P = 0^-$ must be in an orbital angular
momentum $L = 1$ state.  Hence, the discontinuity is expected to
appear in the second derivative of ${\rm Re}\Pi^{VP}(q^2,m_V,m_P)$ at
$q^2 =-(m_{\omega}+m_{\pi})^2$.  Numerical differentiation of the
results shown in \Fig{Fig:VVP} confirm the presence of this
discontinuity.

One observes that $\Pi^{VP}(q^2,m_V,m_P)$ passes through zero at $q^2
= 0$, a feature that is not observed for $\Pi^{PP}(q^2,m_P,m_P)$.
This zero is a result of the structure of the $VVP$ meson transition
amplitude $\Lambda^{ij}_{\mu\nu}(p_1,p_2)$ given in \Eq{VVPvertex},
which is required by Lorentz covariance.  The requirement that this
amplitude be proportional to
\begin{equation}
\epsilon_{\mu\nu\alpha\beta} \; p_{1\alpha}\; p_{2\beta} = 
\epsilon_{\mu\nu\alpha\beta} \;  q_{\alpha} \; p_{1\beta} 
,
\end{equation}
leads to a vector-meson self energy $\Pi_{\mu\nu}^{VP}(q^2,m_V,m_P)$
that is necessarily transverse to $q_{\mu}$ for all values of $q^2$
and hence is of the form given in \Eq{Pimunu}.  It follows that the
vector-meson self energy in this channel vanishes at $q^2 = 0$.

Because the real part of $\Pi_{\mu\nu}^{VP}(q^2,m_V,m_P)$ goes through
zero at $q^2 = 0$, approaches zero in the deep spacelike region and is
infinitely differentiable below the breakup threshold at $q^2 =
-(m_V+m_P)^2$, it is a small and slowly varying function for $q^2 \geq
-m_{\rho}^2$.  Therefore, the $\omega\pi^0$ intermediate state is
expected to contribute very little to the $\rho$ meson self energy
(from \Fig{Fig:VVP}, it is clear that the $K^* K$ state contributes
even less). However, this channel is important for the self energy of
the $\omega$ meson, owing to the fact that there are three different
$\rho\pi$ channels that contribute to the $\omega$-meson self energy;
they are $\rho^+\pi^-$, $\rho^-\pi^+$, and $\rho^0\pi^0$.  This leads
to the overall factor of three in front of
$\Pi_{\mu\nu}^{VP}(q^2,m_{\rho},m_{\pi})$ in \Eq{PiSum:Omega}, which
makes the $\rho\pi$ channel as important for the $\omega$-meson self
energy as the $\pi\pi$ channel is for the $\rho$-meson self energy
(see \Tab{Tab:Shifts}).

\begin{figure}[t]
 \begin{center}
 \epsfig{figure=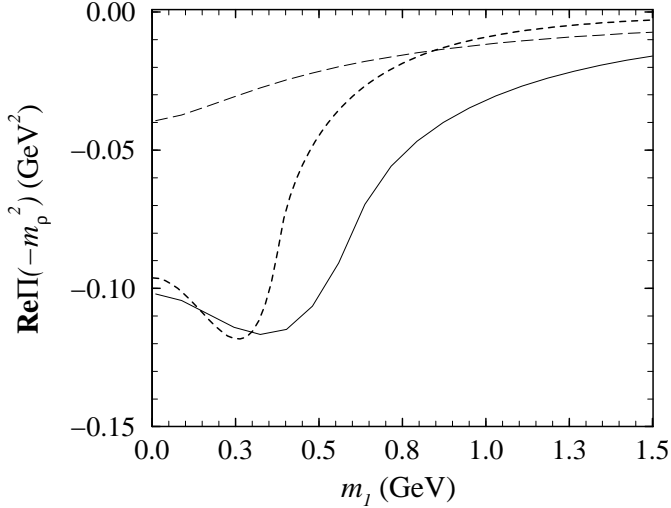,height=7.0cm}
 \end{center}
\caption{
The dependence of the real part of the vector-meson self energy
on the mass $m_1$ of the intermediate state mesons.
The dashed curve shows the dependence of the vector-pseudoscalar channel
$\Pi^{VP}(-m_{\rho}^2,m_\rho,m_1)$
when the mass $m_1$ of the pseudoscalar meson is varied.
The solid curve shows the dependence of the two-pseudoscalar channel
$\Pi^{PP}(-m_{\rho}^2,m_{\pi},m_1)$ 
when the mass $m_1$ of {\em one} of the pseudoscalar mesons is varied.
The short-dashed curve shows the dependence of the two-pseudoscalar
channel $\Pi^{PP}(-m_{\rho}^2,m_1,m_1)$ 
when the mass $m_1$ of {\em both} pseudoscalar mesons is varied.
}
\label{Fig:m1dep}
\end{figure}

A single coupling constant and meson-transition form factors has been
used for each of the two types of intermediate states ($PP$ and $VP$)
considered herein.  The result is that the only difference between the
various channels within a particular type of meson loop is the
masses of the intermediate mesons involved.  A comparison of the self
energies resulting from the $KK$ and $K^*K$ channels with $\pi\pi$ and
$\omega\pi$ channels, respectively, suggests that states with heavier
mesons contribute less than states of lighter mesons.  For example,
for all $q^2 \geq - m_{\rho}^2$, the value of $\Pi^{PP}(q^2,m_K,m_K)$
is less than half the value of $\Pi^{PP}(q^2,m_{\pi},m_{\pi})$.  The
same is true when $\Pi^{VP}(q^2,m_{K^*},m_K)$ is compared with
$\Pi^{PP}(q^2,m_{\omega},m_{\pi})$ in \Fig{Fig:VVP}.

Within the present model, the dependence of the vector-meson self
energy on the mass of the intermediate mesons can be explored.  This
is done by letting one or both of the masses $m_P$ in \Eqs{PiPP} and
(\ref{PiVP}) vary while keeping $q^2 = -m_{\rho}^2$.  The solid curve
in \Fig{Fig:m1dep} is the result of a calculation of the real part of
the self energy $\Pi^{PP}(-m_{\rho}^2,m_{\pi},m_1)$ in which one of
the pseudoscalar-meson propagators in \Eq{PiPP} has mass $m_{\pi}$ and
the other propagator has a mass $m_1$ which is varied from 0 to
1500~MeV.  It is clear from \Fig{Fig:m1dep} that the magnitude of the
self energy depends critically on the mass $m_1$.  In particular, for
$m_1 = m_{\rho}$, the real part of the self energy
$\Pi^{PP}(-m_{\rho}^2,m_{\pi},m_1)$ has already decreased by a factor
of two from its value for $m_1 = m_{\pi}$.  The short-dashed curve
in \Fig{Fig:m1dep} is the self energy $\Pi^{PP}(-m_{\rho}^2,m_1,m_1)$
obtained with the masses of the two pseudoscalar mesons are taken to
be {\em equal} and varied together.  Increasing both masses of the
intermediate mesons at the same time causes the self energy to
decrease very rapidly.

It was argued in \Ref{LC} that lattice calculations which
linearly extrapolate the $\rho$ mass as a function of $m_\pi$ should
receive negligible corrections from the two-pion intermediate
state. The calculations described here are in agreement with
this conclusion, but as can be seen from the short-dashed curve in
\Fig{Fig:m1dep}, the dependence of the real part of the vector
self-energy in the region of the physical pion mass is not smooth,
showing the effects of the threshold crossed at $4\,m_\pi^2$. This
rapid dependence on the mass of the intermediate states suggests that
such extrapolations must be made with caution.

The dashed curve in \Fig{Fig:m1dep} shows the dependence of the self
energy $\Pi^{VP}(-m_{\rho}^2,m_{\omega},m_1)$ when the mass of the
intermediate vector meson is $m_V$ and the mass of the intermediate
pseudoscalar meson $m_1$ is varied.  Because of the different
Lorentz-covariant structure of the vector-vector-pseudoscalar
transition amplitude $\Lambda_{\mu\nu}^{ij}(p_1,p_2)$ given in
\Eq{VVPvertex}, and that the threshold for $\rho \rightarrow \omega
\pi$ has not been reached even for $m_1 = 0$, a much weaker dependence
on the mass $m_1$ is observed compared to that of
$\Pi^{PP}(-m_{\rho}^2,m_1,m_2)$. Nonetheless, it may be sufficiently
accurate to neglect pseudoscalar-vector intermediate states with
masses $m_1 \geq m_{\rho}$ when calculating the self energies, since
these are smaller by at least a factor of two than the contribution
from the $\omega\pi$ intermediate state.

It is interesting to note that the curves in \Fig{Fig:m1dep} all
exhibit an extremum for values of $m_1$ for which the sum of the
intermediate-mesons masses $(m_1 + m_2)$ is within the range
$\sfrac{1}{2} m_{\rho} \leq (m_1 + m_2) \leq m_{\rho}$.  This is not
coincidental, since $m_{\rho}$ is the value of the
center-of-momentum energy at the decay threshold.  
For values of $m_1 > m_{\rho} - m_2$ 
the real part of the amplitude is a monotonic, decaying function of
$m_1$.  At threshold $m_1 = m_{\rho} - m_2$, discontinuities in the
second derivative of the real part of the self energy cause the
rapid turnover observed in \Fig{Fig:m1dep}, so that the extremum of
${\rm Re }\Pi(-m^2_{\rho},m_1,m_2)$ occurs for a value of $m_1$
slightly below the threshold value $m_1 = m_{\rho} - m_2$.

This suggests that in calculations of self energies for a particle of
mass $M$, it may be sufficiently accurate to consider only those
states for which the sum of the intermediate hadron masses $m_1 + m_2
+ \cdots$ lie on the range $\sfrac{1}{2} M \leq \sum_{i} m_i \leq M$.
This implies greatly simplified calculations of mixings between states
where it is only necessary to consider the contributions from
multiple-hadron states within this range of total masses. This is a
general feature of such calculations and not a specific result of the
particular channels considered herein, since it arises independently
of the particular form and scales used to model the meson transition
form factors $f^{VPP}(p_1,p_2)$ and $f^{VVP}(p_1,p_2)$.

\subsection{vector-meson mass shifts}
\label{Sec:ResultsMassShifts}
%
\begin{table}[t]
 \caption
 {Mass shifts to $\rho$ and $\omega$ mesons arising from several 
 two-meson intermediate states. Each channel provides less than a $10\%$ 
 correction to the total mass of the vector meson. 
 The sum of these shifts, which is not observable, is given in the bottom
 row. The {\em difference} between these sums is 
 $m_{\omega} - m_{\rho} =$ 24.8~MeV, which can be compared to the
 experimental value of $12\pm 1$ MeV.   
 }
 \begin{tabular}{c|rr}
 channel & $\Delta m_{\rho}$ (MeV)& $\Delta m_{\omega}$ (MeV)\\ \hline
 $\pi\pi$ 	& -69.8 &       \\
 $K \bar{K} $ 	& -14.8 & -14.8 \\
 $\omega\pi$ 	& -22.5 &       \\ 
 $\rho\pi$ 	&       & -67.5 \\
 $\omega\eta$ 	&       & -13.1 \\ 
 $\rho\eta$ 	& -13.1 &       \\ 
 $K^* K $ 	& -11.9 & -11.9 \\ \hline
 sum 	 	&-132.1 &-107.3 \\ 
 \end{tabular}
 \label{Tab:Shifts}
\end{table}

Results for the mass shifts due to the two-meson loops
considered in this study are summarized in \Tab{Tab:Shifts}.  It
should be noted that only the difference between the total mass shifts
of the $\rho$ and $\omega$ mesons is physically meaningful.  The
absolute size of each term does, however, give some indication of the
importance of the dressing by two-meson loops.  The largest
mass shifts are that of the $\rho$ mass due to the two-pion loop,
and that of the $\omega$ due to the $\rho\pi$ loop, which are about
$-10\%$ of the bare mass. 

Note that the $\rho$ mass shift due to the two-pion loop
is roughly $-70$ MeV, while the two-pion $\rho$ mass shift in \Ref{LC}
was between $-10$ and $-20$ MeV. The difference is due to an
additional contribution to the vector self-energy which arises from a
$\pi\pi\rho\rho$ contact term.  This term provides a contribution to
the vector-meson self energy $\Pi_{\mu\nu}(q)$ that is independent of
$q$.  In \Ref{LC}, this term is added to provide current conservation
for the vector $\rho$; i.e., that $\Pi^{PP}(q^2=0)=0$.  Since one can
show that this term contributes equally to the self energies of the $\rho$
and $\omega$ mesons, it cannot contribute to their mass splitting and
so has not been included here.

To allow for the direct comparison with \Ref{LC}, the effects of
including this contact term into the present study can be reproduced
by subtracting $\Pi(q^2=0)$ from the self energy $\Pi(q^2)$.  This
procedure ensures that the condition obtained in \Ref{LC} that
$\Pi^{PP}(q^2=0)=0$ is reproduced.  The resulting mass shift due to
the two-pion loop obtained from \Eq{massShifts} would then be
approximately 
\begin{eqnarray}
\delta m_{\pi\pi} & \approx &
{\rm Re}\left[
\frac{
\Pi^{PP}(\-m_\rho^2,m_\pi,m_\pi) 
    \- \Pi^{PP}(0,m_\pi,m_\pi)
}{2 \; m_0}\right]
\nn\\
& \approx & - 9.2\ {\rm MeV}.
\end{eqnarray}
This mass shift for the $\rho$ meson is similar to that of \Ref{LC}.

As pointed out in \Ref{GI}, the strange-meson and vector-$\eta$
intermediate states cannot contribute to the mass splitting; they are
included here to illustrate that as the masses of the intermediate
state mesons increase, the resulting shifts decrease rapidly. For
example, the shift in the $\rho$ mass due to the $K\bar{K}$
intermediate state is about one fifth of that due to two pions.
Similarly, the shift in the $\omega$ mass due to the $\rho\eta$
intermediate state is roughly half the size of that due to
$\omega\pi$. 

The result of adding the shifts due to the various
intermediate states is that they largely cancel, as in \Ref{GI}, to
give a mass splitting of roughly 25 MeV, which is of the same order
(on the scale of the bare mass) as the experimental value of $12\pm 1$
MeV. Of course, a complete calculation would have to consider all
possible multiple-hadron intermediate states. It is encouraging that,
within this model, a reasonable value of the mass splitting is
achieved using only a handful of meson loops, and that the
individual contributions due to additional loops are expected to be
small compared to the dominant contributions evaluated here.

\subsection{time orderings}
\label{Sec:ResultsOrders}
%
%
\begin{figure}[t]
\begin{center}
\epsfig{figure=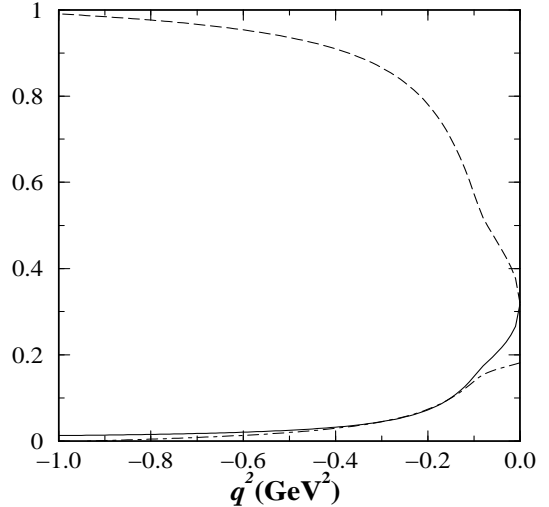,height=7.0cm,width=7.0cm}
\end{center}
\caption{ Ratios of the forward-forward (dashed curve),
backward-backward (solid curve) and forward-backward or
backward-forward (short-dashed long-dashed curve) time orderings of
the real part of the two-pion loop integral to
the sum of all four, as a function of the vector-meson mass
squared.}
\label{Fig:Ratios}
\end{figure}

Figure~\ref{Fig:Ratios} shows the ratio of the forward-forward
$\Pi^{[++]}(q^2)$, backward-backward $\Pi^{[--]}(q^2)$, and
forward-backward $\Pi^{[+-]}(q^2)$ [equal to backward-forward
$\Pi^{[-+]}(q^2)$] time orderings of the real part of the two-pion
loop integral to their sum $\Pi^{\pi\pi}(q^2)$, as a function of the
vector-meson mass squared $q^2$. As expected from the analysis in
Sec.~\ref{Sec:Orders}, at $q^2=0$ the forward-forward time ordering
$\Pi^{[++]}(q^2)$ is equal to the backward-backward time ordering
$\Pi^{[--]}(q^2)$. It is also simple to show [by a change of variables
from $k_4$ to $-k_4$ in Eq.~\ref{TOPiPP}] that
$\Pi^{[+-]}(q^2)=\Pi^{[-+]}(q^2)$ at all values of $q^2$.
Although not shown here, an examination of the
derivative of $\Pi^{[++]}(q^2)$ indicates a discontinuity at
$q^2=4m_\pi^2$ indicating the onset of a branch cut in
$\Pi^{\pi\pi}(q^2)$ due to $\Pi^{[++]}(q^2)$, which verifies the
identification of this time ordering with the propagation of two
forward-propagating pions.

In Fig.~\ref{Fig:Ratios}, one observes that at $q^2 = - m_{\rho}^2$ the
forward-forward time ordering is dominant and the backward-backward
contribution is only a few percent of the total. This realizes the
expected strong suppression of the backward-backward time ordering,
due to the vertices and the large energy denominators for the
$\rho\rho\pi\pi$ intermediate state. Note also that both the
forward-forward and backward-backward contributions have the same
sign, so that the two-pion loop contribution to the
$\rho$ self energy is not reduced by a cancellation between these two
time orderings. This establishes that, in the present quantum
field-theoretic framework, the dominant contribution to the $\rho$
self energy is from two forward-propagating pions, and that the
difference between the mass shifts obtained in this approach and those
of the quantum-mechanical approach of Ref.~\cite{GI} is not due to the
presence of additional time orderings in a Lorentz covariant framework.

As discussed in Sec.~\ref{TOPiPP}, the forward-backward and
backward-forward time orderings are not zero in the present framework;
Fig.~\ref{Fig:Ratios} shows that they also make negligible
contributions to $\Pi^{\pi\pi}(q^2)$ at $q^2=-m_\rho^2$. 

\subsection{vector-meson electromagnetic form factors}
\label{Sec:ResultsEMffs}
%
\begin{figure}[t]
\begin{center}
\epsfig{figure=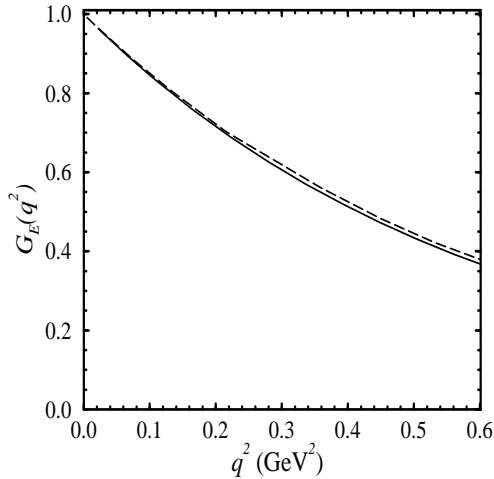,height=7.0cm,width=7.0cm}
\end{center}
\caption{ The $\rho$-meson electric-monopole form factor.  The
dashed curve is the result $G_{E}^{\bar{q}q}(q^2)$ from \Ref{Hawes}
that rises from quark-antiquark substructure of $\rho$ meson.  The
solid curve is the real part of the form factor $G_{E}(q^2)$ that
results when the effects of pion loops from \Eq{Gpipik2} are added to
the results from \Ref{Hawes}.  }
\label{Fig:Ge}
\end{figure}

The final application of vector-meson dressing by two-meson loops
that is considered here is the correction to the EM form factors and
charge radius of the $\rho$ meson.  In \Sec{Sec:EMFF}, an expression
for the pion contribution to the EM form factor $G_{i}^{\pi\pi}(Q^2)$
was obtained in terms of the same meson transition form factor
$f^{VPP}(p_1,p_2)$, given in \Eq{fVPPForm} with \Eq{VPPFit}, that was
used to obtain the $\pi\pi$ and $K\bar{K}$ contributions to the
$\rho$-meson self energy.

In the present calculation of $G^{\pi\pi}_E(Q^2)$, the final term in
\Eq{Gpipik2} (arising from the surface contribution of the integration
by parts) is small and varies slowly with $Q^2 \approx 0$, hence it is
safe to neglect this term for the determination of the charge radius.  
The resulting $G_{E}^{\pi\pi}(Q^2)$ is well described for spacelike values
of the photon momentum squared $0 < Q^2 < $ 0.8~GeV$^2$ by the form
\begin{equation}
G_{E}^{\pi\pi}(Q^2) \approx A \; e^{-B Q^2 / m_{\rho}^2}
\end{equation}
with $B \approx$ 1.39 and the value of $A$ as determined from 
\Eq{deltaZpipi} is $A = \delta Z_{\pi\pi} = $ 0.134.

The form factor $G_{E}^{\pi\pi}(Q^2)$ is added to the quark-antiquark
contribution $G_{E}^{q\bar{q}}(Q^2)$ obtained by \Ref{Hawes} according to
\Eq{Ge}.  The resulting form factor $G_{E}^{\pi\pi+q\bar{q}}(Q^2)$ is
plotted as a solid curve in \Fig{Fig:Ge}.   It falls more rapidly with the
photon momentum $Q^2$ than 
does the form factor $G_{E}^{q\bar{q}}(Q^2)$ (dashed curve) indicating
that the inclusion of the $\pi\pi$ intermediate state into the $\rho$
meson self energy has {\em increased} the charge radius of the $\rho$
meson.  The charge radius of the $\rho$ meson due to its quark-antiquark
substructure and its mixing with the two-pion state obtained using 
\Eq{RadiusDef} is
\begin{equation}
{\langle r^2_{\bar{q}q + \pi\pi} \rangle }^{\sfrac{1}{2}} 
	 =  0.67 \;\;{\rm fm}.
\end{equation}
This value may be compared to the value obtained in \Ref{Hawes}, 
\begin{equation}
{\langle r^2_{\bar{q}q} \rangle }^{\sfrac{1}{2}} 
	 =  0.61 \;\;{\rm fm}
, 
\end{equation}
which results from only the quark-antiquark substructure of the $\rho$
meson. One observes that including pion loops increases 
the charge radius of the $\rho$ meson by approximately $10\%$.
This is consistent with the increase of less than $15\%$ observed in a similar
study of the pion-loop contribution to the $\pi$ meson charge radius
\cite{Bender}. 

Although, the short lifetime of the $\rho$ meson excludes the possibility
of directly measuring the EM charge radius of the $\rho$ meson at present,
one can define a ``diffractive radius'' from the $t$ dependence of
diffractive $\rho$-meson electroproduction cross sections \cite{MAP}. 
As with the EM charge radius, the diffractive radius receives
contributions from its quark substructure as well as from mixings with
multiple-hadron states.  
However, an important distinction between the charge radius and the
diffractive radius is that since diffraction arises from a strong
interaction, electromagnetically neutral particles, such as the $\omega$
and $\rho^0$ mesons,  will have diffractive radii even though they have no
well-defined charge radius\footnote 
{
To be electroproduced diffractively, the vector meson must be able to mix
with the photon, and so must be neutral (e.g., $\omega$,
$\rho^0$, $\phi$, $J/\psi$).
Therefore, {\em only} for neutral vector mesons can one define a
diffractive radius. 
}.

The results of this study suggest that the most significant
contribution from two-meson loops to the diffractive
radius of the $\rho$ meson would be due to the two-pion loop.
However, $G$-parity forbids the $\omega$ meson from mixing with the
two pions, so that the diffractive radius of the $\omega$ meson
would be unaffected by the two-pion loop.  Therefore, the diffractive
radius of the $\omega$ meson is expected to be {\em smaller} than that
of the $\rho$ meson.  An observed difference in the $t$-dependence of
diffractive electroproduction of $\rho$ and $\omega$ mesons would be
indicative of their differing diffractive radii, and such a
measurement could provide a means to estimate the contributions to
these radii from pion loops.

\section{Conclusions}
\label{Sec:Concl}

It is shown that, in a covariant model based on studies of the
Schwinger-Dyson equation of QCD which assumes exact SU(3)-flavor
symmetry, the contributions to the self energies of the $\rho$ and
$\omega$ due to several pseudoscalar-pseudoscalar and
pseudoscalar-vector meson loops are at most $10\%$ of the bare
mass. The result for the mass shift of the $\rho$ meson due to the
two-pion loop is in agreement with a previous
calculation of this quantity using an effective chiral Lagrangian
approach. Such contributions are found to decrease rapidly as the mass
of the intermediate mesons increases beyond $m_{\rho}/2$.

The mass shifts due to several two-meson intermediate states are
compared with those from an extensive study within a nonrelativistic
framework of the $\rho$-$\omega$ mass splitting, and are found to be
smaller, especially for intermediate states involving higher-mass
mesons. A mass splitting of $m_{\omega} - m_{\rho}\approx 25$~MeV is
found from the $\pi\pi$, $K\bar{K}$, $\omega\pi$, $\rho\pi$,
$\omega\eta$, $\rho\eta$ and $K^* K$ channels. A complete calculation
which evaluates contributions from all possible multiple hadron
intermediate states and breaks SU(3)-flavor symmetry is beyond the
scope of the present work. However, these results suggest that such a
calculation should exhibit rapid convergence as the number of
two-meson intermediate states is increased by including states with
higher masses. This implies that inclusion of two-meson loops
into the vector-meson self-energy yields a small correction to
the predominant valence quark-antiquark structure of the vector meson.

The part of the vector-meson self energy which corresponds to two
pions propagating forward in time has been shown to dominate a
Lorentz-covariant calculation of the two-pion loop contribution. This
implies that the additional time-orderings of this loop, necessary to
maintain Lorentz covariance, are not responsible for the reduced size
of the mass shifts found in the present framework when compared to
quantum-mechanical treatments.

As a test of the self-consistency of this calculation, the
contribution of the two-pion loop to the $\rho$ meson EM
form factor is evaluated, and is shown to provide a modest increase of
about $10\%$ to the charge radius of the $\rho$. Although this charge
radius is difficult to observe, it may be possible to observe the
effects of such an increase by examining the difference in the
``diffractive radii'' of the neutral $\rho$ and $\omega$ found from
the $t$-dependence of diffractive electroproduction cross sections. As
the two-pion loop cannot contribute to the $\omega$, it may
be possible to observe its effects on the $\rho$ by measuring a
difference in these diffractive radii.

\section{Acknowledgments}
The authors wish to thank A.~Szczepaniac for illuminating discussions.
This work is supported by the U.S. Department of Energy
under contracts DE-AC05-84ER40150, DE-FG05-92ER40750 and 
DE-FG02-87ER40365, 
and the Florida State University Supercomputer Computations Research
Institute which is partially funded by the Department of Energy through 
contract DE-FC05-85ER25000.


\begin{references}
\bibitem{MT} K.~L.~Mitchell and P.~C.~Tandy, 
		Phys.\ Rev.\ C {\bf 55}, 1477 (1995).
\bibitem{LC} D.~B.~Leinweber and T.~D.~Cohen, 
		Phys.\ Rev.\ D {\bf 49}, 3512 (1994).
\bibitem{GI} P.~Geiger and N.~Isgur, 
		Phys.~Rev.~Lett.~{\bf 67}, 1066 (1991); 
		Phys.~Rev.~D {\bf 41}, 1595 (1990); 
		{\it ibid}~D {\bf 44}, 799 (1991).
\bibitem{Hollenberg} L.~C.~L.~Hollenberg, C.~D.~Roberts and 
	B.~H.~J.~McKellar, Phys.\ Rev.\ C {\bf 46}, 2057 (1992).
\bibitem{Hawes} F.~T.~Hawes and M.~A.~Pichowsky, 
		Phys.\ Rev.\ D {\bf 59}, 1743 (1999).
\bibitem{Bender} R.~Alkofer, A.~Bender and C.~D.~Roberts,
	Int.\ J.\ Mod.\ Phys.\ A {\bf 10}, 3319 (1995).
\bibitem{MAP} M.~A.~Pichowsky, ``Diffractive radii of hadrons,''
	in preparation.
\bibitem{Burden}  C.~J.~Burden, C.~D.~Roberts and M.~J.~Thomson,
    		Phys.\ Lett.\ B {\bf 371}, 163 (1996);\\
		C.~D.~Roberts, Nucl.\ Phys.\ A {\bf 605}, 475 (1996).
\bibitem{NewOnes} 
	M.~A.~Ivanov, Yu.~L.~Kalinovsky, P.~Maris, and C.~D.~Roberts,
	Phys.\ Rev.\ C {\bf 57}, 1991 (1998); \\
	Yu.~Kalinovsky, K.~L.~Mitchell and C.~D.~Roberts,
	Phys.\ Lett.\ B {\bf 399}, 22 (1997); \\
	M.~.A.~Pichowsky and T.-S.~H. Lee, 
	Phys.\ Rev.\ D {\bf 56}, 1644 (1997); \\
	R.~Alkofer, C.~.D.~Roberts, 
	Phys.\ Lett.\ B {\bf 369}, 101 (1996);	\\
	M.~R.~Frank, K.~L.~Mitchell, C.~D.~Roberts, P.~C.~Tandy, 
	Phys.\ Lett.\ B {\bf 359}, 17 (1995).
\bibitem{Frank} M.R.~Frank and C.D.~Roberts,
	Phys.\ Rev.\ C {\bf 53}, 390 (1996).
\bibitem{Maris}	P.~Maris and C.D.~Roberts,
	Phys.\ Rev.\ C {\bf 56}, 3369 (1997); \\
	P.~Maris and C.D.~Roberts, 
	Phys.\ Rev.\ C {\bf 58}, 3659 (1998).
\bibitem{Wightman} 
	C.~D.~Roberts and A.~G.~Williams, 
	Prog.\ Part.\ Nucl.\ Phys.\ {\bf 33}, 477 (1994), 
	and references therein. 
\bibitem{Roberts} C.~D.~Roberts, 
	``Nonperturbative QCD with modern tools'', lectures given at the 
	11th Physics Summer School on Frontiers in Nuclear Physics,
	Australia National University, Canberra, Australia, 
	LANL e-print \# nucl-th/9807026, 1988.
\bibitem{BQRTT} C.~J.~Burden, L~.Qian, C.~D.~Roberts, P.~C.~Tandy and
	M.~J.~Thomson, Phys.\ Rev.\ C {\bf 55}, 2649 (1997).
\bibitem{MAP2} The importance of including higher Dirac moments in decay
	processes like $\rho\rightarrow \pi\pi$ was suggested to us by
	E.~S.~Swanson, based on quark-model studies of meson decays 
	[see P.~Geiger and E.~S.~Swanson, Phys.\ Rev.\ D{\bf 50}, 6855
	(1994)]. 
	It has since been verified, in calculation of the decay
	$\rho\rightarrow\pi\pi$ using the model of \Ref{BQRTT}, 
	that discarding all but the term with the largest strength 
	(proportional to $\gamma_5$) in the pion 
	Bethe-Salpeter amplitude tends to increase the coupling constant
	$g_{\rho\pi\pi}$ by $71\%$ 
	[see P.C.~Tandy, ``Modeling nonperturbative QCD for mesons and
	couplings'', Proceedings of the IVth Workshop on 
	Quantum Chromodynamics, Paris (1998) 
	LANL e-print \#nucl-th/9812005].
\bibitem{MAP3} In contrast to decays of a vector meson into two
	pseudoscalar mesons, the transition amplitude of a vector-meson
	coupling to a vector and pseudoscalar meson is rather insensitive
	to sub-leading Dirac amplitudes in the meson Bethe-Salpeter
	amplitudes.   Hence, the value for the coupling constant
	$g_{\omega\rho\pi}$ obtained herein, is more reliable than the
	result for $g_{\rho\pi\pi}$ obtained in \Ref{Hawes}.
\bibitem{UMeissner} V.~Bernard, N.~Kaiser and U.~G.~Meissner, 
	Eur.\ Phys.\ J.\ A {\bf 4}, 259 (1999).
\bibitem{Durso} J.~W.~Durso, Phys.\ Lett.\ B {\bf 184}, 348 (1987).
\end{references}
\end{document}